\documentclass[12pt]{iopart}

\usepackage[dvips]{graphicx}%
\usepackage{bm}
\usepackage{cite}
\usepackage{multirow}
\usepackage[normalem]{ulem} 
\usepackage[colorlinks=true,linkcolor=blue,citecolor=blue,urlcolor=blue]{hyperref}

\newcommand{\op}[1]{{\bm{#1}}}
\newcommand{\bra}{\langle}
\newcommand{\ket}{\rangle}

\newcommand{\Oplus}{\ensuremath{\vcenter{\hbox{\scalebox{1.5}{$\oplus$}}}}}

\begin{document}

\title[Quantum Phase Diagrams of Matter-Field Hamiltonians]{Quantum Phase Diagrams of Matter-Field Hamiltonians I:\\
Fidelity, Bures Distance, and Entanglement}
\author{Sergio Cordero, Eduardo Nahmad-Achar, Ram\'on L\'opez-Pe\~na, Octavio Casta\~nos} 

\address{
Instituto de Ciencias Nucleares, Universidad Nacional Aut\'onoma de M\'exico, Apartado Postal 70-543, 04510 Cd. Mx., Mexico }

\ead{sergio.cordero@nucleares.unam.mx}

\date{\today}

\begin{abstract}
A general procedure is established to calculate the quantum phase diagrams for finite matter-field Hamiltonian models. The minimum energy surface associated to the different symmetries of the model is calculated as a function of the matter-field coupling strengths. By means of the ground state wave functions, one looks for minimal fidelity or maximal Bures distance surfaces in terms of the parameters, and from them the critical regions of those surfaces characterize the finite quantum phase transitions. Following this procedure for $N_a=1$ and $N_a=4$ particles, the quantum phase diagrams are calculated for the generalised Tavis-Cummings and Dicke models of 3-level systems interacting dipolarly with $2$ modes of electromagnetic field.
For $N_a=1$, the reduced density matrix of the matter allows us to determine the phase regions in a $2$-simplex (associated to a general three dimensional density matrix), on the different $3$-level atomic configurations, together with a measurement of the quantum correlations between the matter and field sectors. As the occupation probabilities can be measured experimentally, the existence of a quantum phase diagram for a finite system can be established.
\end{abstract}

%
%


\section{Introduction}
The subject of continuous quantum phase transitions is of interest because it appears in many-body systems, has been studied in nuclear, molecular, quantum optics, and condensed matter physics, all of which have potential applications in the design of quantum technologies. A qualitative account of the quantum phase transitions is given in~\cite{sondhi97}. These transitions take place at a temperature of absolute zero, and at the critical points of a parameter in the Hamiltonian where the nature of the ground state changes in a fundamental way~\cite{sondhi97,sachdev11}.

The analysis of the phase diagram structure of a system of atoms in the presence of a radiation field in a QED cavity is important, mainly due to the presence of quantum phase transitions. It is characteristic in these diagrams the distinction of a {\it normal} and a {\it collective} regime, the difference being the decay rate: for any two-level subsystem, it is linear in the number of particles $N_a$ in the normal region (the expected result for independent atomic emission); whereas in the collective (also called {\it superradiant}) region, it is proportional to $N_a(N_a + 2)$, which for a large number of particles goes like $N_a^2$~\cite{hepp73, nussenzveig73}. By analogy, we will call here a subregion of the phase diagram as {\it normal} when the decay rate is proportional to $N_a$, and {\it collective} or {\it superradiant} when proportional to $N_a(N_a + 2)$.

Being quantum optical systems the foundational elements for quantum information and quantum computing, the interest in these studies goes beyond the mere understanding of how they behave into the realm of serious applications, with particular interest in the case of a {\it finite} number of particles. Furthermore, systems with a finite number of levels constitute important simplified models for the interaction between matter and radiation which are not only physically realistic, but tractable mathematically~\cite{einstein17}. A treatment of the early work on cavity QED with quantised modes demonstrating the reaction of the matter back to the field may be found in~\cite{yoo85}. There, the interaction of a $2$- or $3$-level atom interacting with quantised cavity fields in the rotating wave approximation was studied using the time-dependent Schr\"odinger equation.

The quantum phase diagrams for three-level atoms interacting with a single mode of the electromagnetic field have been determined, in the limit $N \to \infty$, by means of the Holstein-Primakoff boson mapping, including the diamagnetic term. The inequalities imposed by the Thomas-Reich-Kuhn oscillator strength sum rules are preserved~\cite{baksic13}.  For the quantum Rabi and Jaynes-Cummings models, a Universal critical behaviour has been established which is valid for a finite number of particles~\cite{liu17, wei18}. (The quantum Rabi model is the particular case of the Dicke model for a single atom; see the special volume dedicated to a review of the semi-classical and quantum Rabi model~\cite{braak16}.) The finite size critical behaviour of the adiabatic Dicke model has been studied in~\cite{liberti10}, which exhibits many features in common with the uniaxial model for spin system.

The characterisation of quantum phase transitions in terms of the overlap function (fidelity) between two ground states was  first described in~\cite{zanardi06}, and the procedure was applied to several Hamiltonian models. The notion of fidelity has been used in the context of information theory as a measure of distinguishability of probability distributions, and applications to the study of phase transitions in condensed matter systems and some of its geometric features have been presented in~\cite{vieira10}. For a pedagogical review of the quantum fidelity approach to quantum phase transitions see~\cite{gu10}.

In the area of quantum critical phenomena, the importance of quantum information concepts like the fidelity and the susceptibility of the fidelity has been growing in the last decade, mainly because they measure the similarity between states and we know that at critical points there is a strong change in the nature of the ground state. A relation between the fidelity and the dynamic structure factor for the driven Hamiltonian has been established, which allows to calculate the fidelity by means of several numerical techniques~\cite{you07}. These concepts have also been used to extract the quantum criticality and to do state engineering in a simulated anisotropic Rabi Hamiltonian by means of a circuit QED system with superconducting flux qubits~\cite{wang18}.

Properties of the phase transitions appearing in the Dicke, Tavis-Cummings, Rabi, and Jaynes-Cummings models have been analysed in~\cite{larson17} in which, through an adiabatic approximation, they obtain a mean-field description, neglecting quantum fluctuations. These approximations do not, however, satisfy the symmetries of the Hamiltonian, which may be restored through projected or symmetry-adapted states~\cite{castanos11, castanos09b}.

In this work we study finite systems, and see that strong changes in the composition of the ground state (and even in the first excited states) take place when we cross singular regions in the energy surface. The phase diagram in this finite case is much richer than that obtained in the thermodynamic limit, in that many more transitions appear. As the number of particles $N_a$ increases, these new transitions deform into those that remain in the thermodynamic limit.We have therefore called these ``phase transitions'', justified by the latter argument ({\it vide infra}). The {\it finite} quantum phase diagram is corroborated by the results of the quantum correlations between the matter and field sectors, and by the calculation of the occupation probabilities. A finite-effect characterisation of the quantum phase transitions in terms of their continuity and stability is proposed.

In~\cite{cordero19} we obtained the phase diagrams for the variational result in all the $3$-level atomic configurations, and the finite result for the $\Lambda$ configuration with one particle. Then in~\cite{cordero19b} we presented optimal bases for the general case of $n$-level systems interacting with $\ell$ field modes. In these works, our aim was study the convergence of the solution for the truncated bases, which approach the exact quantum ground state with a desirable precision via a fidelity criterion, and we presented as examples the ground energy surface and separatrices of the $\Lambda$-configuration for a single atom and $\Xi$-configuration for $N_a = 4$ atoms. Here we {\it use} the optimal bases and study the quantum phase diagram for the {\it finite} case in all $3$ configurations: $\Xi$, $\Lambda$, and $V$. We carry out the construction of the ground state energy surfaces for both the generalised Tavis-Cummings model and the generalised Dicke model Hamiltonians. The full quantum solution is obtained and no approximation is made, except for the truncation in the number of photons in order to bring the Hilbert space to a finite dimension. The quantum phase diagrams for the various atomic configurations are built by using the fidelity or the Bures distance between neighbouring states, and the parity (in terms of the symmetry operators for the system) of the ground states that conform these energy surfaces in different regions is studied. The finite quantum phase diagram is corroborated by the results of the quantum correlations between the matter and field sectors, and by the calculation of the occupation probabilities. A finite-effect characterisation of the quantum phase transitions in terms of their continuity and stability is proposed. As mentioned above, we have chosen to name these transitions as {\it quantum phase transitions}, as opposed to {\it crossovers}, {\it precursors}, or other names sometimes given in the literature, since, physically, the ground state suffers considerable changes in its structure, as seen by the mix of basis states that enter into play, when minute fluctuations in any of the system parameters are had, in the exact quantum solution. Physically, then, the ground state acquires a different face and properties.

The reduced density matrix for the matter sector is studied, where we see that for all the different atomic configurations there are regions in the parameter space of the coupling strengths where only two-level subsystems are affected by the Hamiltonian. By calculating the linear entropy, we show that there are quantum correlations between the matter and the electromagnetic field for most values of the occupation probabilities of the energy levels of the system. A simplex for this reduced density matrix is obtained by calculating the occupation probabilities of the three level atomic systems, which can be measured experimentally and can demonstrate the changes established in the phase diagrams. It also gives a visual information about the entanglement properties.

For ease of understanding, explicit examples are given in the case of $3$-level atoms in all their configurations interacting with $2$ modes of an electromagnetic field, but the work can be easily generalised to atoms of any number of levels and to any number of radiation modes.

This paper is organised as follows: 
Section~\ref{model} describes the Hamiltonian for a system of $N_a$ atoms of $n$ levels under dipolar interaction with $m$ modes of an electromagnetic field, where only one mode promotes transitions between two given atomic levels. The symmetries of the Hamiltonian are determined, as well as the basis states in terms of direct products of harmonic oscillators for the field and matter sectors, and its dimension. The construction of the ground state energy surfaces for both the generalised Tavis-Cummings model (GTCM) and the generalised Dicke model (GDM) Hamiltonians is presented in section~\ref{QPD}, and the quantum phase diagrams for the various atomic configurations of $3$-level systems are built by using the fidelity between neighbouring states. The parity of the ground states that conform these energy surfaces in different regions is also studied. Section~\ref{phase_transitions} deals with the characterisation of the quantum phase transitions described earlier in terms of their continuity and stability properties. The reduced density matrix for the matter sector is studied in section~\ref{matter}, as well as the matter-field correlations through the linear entropy. We end in section~\ref{conclusions} with a summary of the results presented.

\section{Model}
\label{model}

\begin{figure}
\begin{center}
\includegraphics[width=0.65\linewidth]{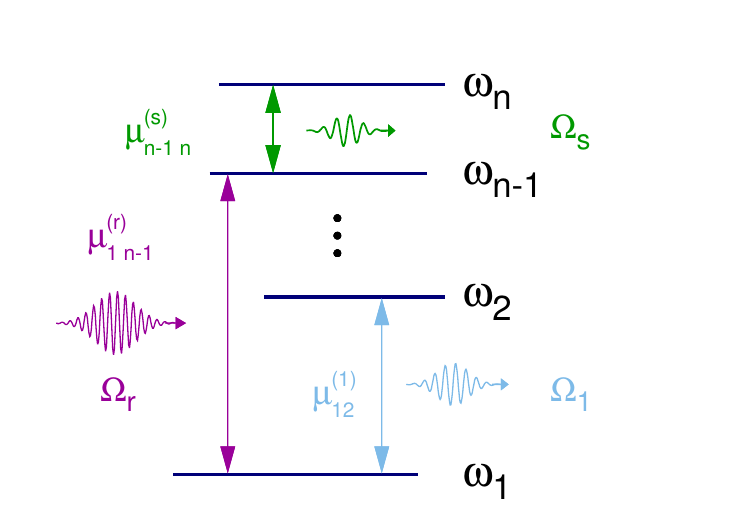}
\end{center}
\caption{Schematics of an $n$-level atom interacting with $\ell$ modes of electromagnetic field. Transitions between two atomic levels $\omega_j\rightleftharpoons \omega_k$ promoted by the mode $\Omega_s$ is indicated with a nonzero value of the dipolar strength $\mu_{jk}^{(s)}$.}\label{nlevel}  
\end{figure}

In this work we study a generalised Dicke model (GDM) by considering a system of $n$-level atoms interacting dipolarly with $\ell$ modes of an electromagnetic field in a cavity. The Hamiltonian for this system is of the form
\begin{equation}\label{eq.H}
\op{H} = \op{H}_D + \sum_{s=1}^\ell \op{H}_{int}^{(s)} \, .
\end{equation}
Here, $\op{H}_D$ is the diagonal part in the harmonic oscillator basis states $\vert \nu_\alpha \, b_k\rangle$ with $\alpha=1,2,\cdots,\ell$ and $k=1,2,\cdots, n$, given by
\begin{eqnarray}\label{eq.HD}
\op{H}_{\textsc{d}} &=& \sum_{s=1}^{\ell} \Omega_{s}\, \op{\nu}_{s}+ \sum_{j=1}^{n} \omega_j \, \op{A}_{jj}\, ,
\end{eqnarray}
where we use $\hbar =1$; $\Omega_{s}$ denotes the field frequency of the $s$-th mode, $\op{\nu}_s=\op{a}_{s}^\dag\, \op{a}_{s}$ the $s$-th photon number operator  with $\op{a}_{s}^\dag,\, \op{a}_{s}$ the corresponding boson creation and annihilation operators, $\omega_j$ the $j$-th atomic level, for which we adopt as convention $\omega_j<\omega_k$ for $j<k$ and fix $\omega_1=0$ and $\omega_n=1$ (so the frequencies and energies are given, respectively, in units of $\omega_n$ and $\hbar\omega_n$), and $\op{A}_{jj}$ is the $j$-th level atomic population operator.

$\op{H}^{(s)}_{\rm int}$ is associated to the dipolar interaction of $s$-mode photons with the $n$-level atoms, 
\begin{eqnarray}
\op{H}_{int}^{(s)} &=&- \frac{1}{\sqrt{N_a}} \sum_{j<k}^{n} \mu_{j k}^{(s)} \left(\op{A}_{jk}+\op{A}_{kj}\right)\left(\op{a}_{s} + \op{a}_{s}^\dag\right)\,. \label{eq.Hint}
\end{eqnarray}
Here $\mu_{jk}^{(s)}$ stands for the dipolar coupling strength and the collective atomic operators $\op{A}_{jk}$ obey the U$(n)$ algebra, i.e.,
\begin{equation}
\left[\op{A}_{lm},\op{A}_{kj}\right] = \delta_{mk}\,\op{A}_{lj}-\delta_{jl}\,\op{A}_{km}\, .\label{eq.commAij} 
\end{equation}
The total number of atoms is determined by the first order Casimir operator
\begin{equation}
\sum_{k=1}^n \op{A}_{kk} = N_a \op{I} \, . \label{eq.Na}
\end{equation} 
This model is simplified by considering that transitions between a pair of atomic levels are promoted only by one mode of the electromagnetic field, which is enforced by the  condition~\cite{cordero16}
\begin{equation}\label{eq.cond1}
\textrm{If}\quad \mu_{jk}^{(s)}\neq 0 \quad \textrm{then}  \quad\mu_{jk}^{(s')}= 0 \quad\textrm{for}\quad s'\neq s\,.
\end{equation}
Notice that this condition implies $1\leq\ell\leq n$ and when $\ell=\ell_0$ with $\ell_0$ the maximum number of permitted atomic transitions, the full system may be divided into two-level sub-systems~\cite{cordero15}.

We should also note that the interaction term (\ref{eq.Hint}) can be divided into two contributions, viz. the rotating and counter-rotating terms; these read for each mode $\Omega_s$ as
\begin{eqnarray}
\op{H}_{R}^{(s)} &=&- \frac{1}{\sqrt{N_a}} \sum_{j<k}^{n} \mu_{j k}^{(s)} \left(\op{A}_{jk}\,\op{a}_{s}^\dag+\op{A}_{kj}\,\op{a}_{s}\right)\,, \\[3mm]
\op{H}_{C}^{(s)} &=&- \frac{1}{\sqrt{N_a}} \sum_{j<k}^{n} \mu_{j k}^{(s)} \left(\op{A}_{jk}\,\op{a}_{s}+\op{A}_{kj}\,\op{a}_{s}^\dag\right)\,, 
\end{eqnarray}
respectively.

Thus, the Hamiltonian in the rotating wave approximation (RWA), namely the generalised Tavis-Cummings model (GTCM), which preserves the total number of excitations, is obtained by neglecting the counter-rotating terms in (\ref{eq.H}), i.e., 
\begin{equation}\label{eq.HRWA}
\op{H}_{\textsc{rwa}} = \op{H}_D + \sum_{s=1}^\ell \op{H}_R^{(s)}\,.
\end{equation} 

For $n$-level atoms interacting dipolarly with $\ell$ modes of an electromagnetic field, it is well known that there are different atomic configurations. Notice that our convention on the atomic levels allows us to determine a particular atomic configuration by choosing the appropriate dipolar strength $\mu_{jk}^{(s)}$ to vanish. As an example of this, the non-zero dipolar strengths for the atomic configurations of $3$-level atoms interacting with two field modes are given in table~\ref{t.3level}. 

\begin{table}
\caption{The non-zero dipolar coupling strengths for $3$-level atoms interacting with two modes of electromagnetic field determine a particular atomic configuration.}\label{t.3level}
\vspace{2mm}
\begin{center}
\begin{tabular}{c |  c}
Configuration & non-zero Dipolar strengths \\[1mm] \hline &\\[-2mm]
$\Xi$ & $\mu_{12}^{(1)},\ \mu_{23}^{(2)}$ \\[2mm]
$\Lambda$ & $\mu_{13}^{(1)},\  \mu_{23}^{(2)}$ \\[2mm]
$V$ & $\mu_{12}^{(1)},\ \mu_{13}^{(2)}$ 
\end{tabular}
\end{center}
\end{table}

We have found that it is very useful to use the critical values of the dipolar couplings for $2$-level systems~\cite{cordero19,cordero19b} 
\begin{equation}
\bar{\mu}_{jk}^{(s)}:=\frac{1}{2}\sqrt{\Omega_s\,\omega_{jk}}; \qquad \omega_{jk}:=|\omega_j-\omega_k|\,,
\end{equation}
and rewrite the Hamiltonian in terms of dimensionless dipolar strengths $x_{jk}^{(s)}$ and detuning parameters $\Delta_{jk}^{(s)}$:
\begin{equation}
x_{jk}^{(s)} := \frac{\mu_{jk}^{(s)}}{\bar{\mu}_{jk}^{(s)}}\,, \qquad \Delta_{jk}^{(s)}:=\frac{\Omega_s}{\omega_{jk}}-1\,.
\end{equation}

We should stress that the ground state energy surface of any $2$-level system only depends on the ratio between the frequency of the field $\Omega$ and the atomic level separations.

\subsection{Symmetries}
\label{symmetries}

The Hamiltonian~(\ref{eq.H}) preserves the number of particles $N_a$, and by simple inspection one notes that at least the parity of the total number of excitations is preserved. In order to find the preserved parities of the full Hamiltonian~(\ref{eq.H}) it is convenient to consider the case of the rotating wave approximation~(\ref{eq.HRWA}), because each constant of motion of~(\ref{eq.HRWA}) is related to a parity symmetry of the full Hamiltonian~(\ref{eq.H}). Thus, we propose as a general constant of motion the linear operator 
\begin{equation}\label{eq.opK}
\op{K} = \sum_{s}^\ell \eta_{s}\, \op{\nu}_{s} + \sum_{j=1}^n \lambda_j\,\op{A}_{jj}\,,
\end{equation}
where the coefficients $\eta_s$ and $\lambda_j$ are obtained by imposing the condition $[\op{H}_{\textsc{rwa}},\op{K}]=0$. This yields a system of $R$ independent linear equations given by
\begin{equation}\label{eq.rank}
\mu_{jk}^{(s)}(\eta_{s} + \lambda_j - \lambda_k) =0\,,
\end{equation}
where $R$ is the rank of the system. We then have $\zeta_0=\ell+n-R-1$ independent constants of motion, apart from the total number of particles $N_a$. As an example, table~\ref{t.opK} shows the operators $\op{K}$ for each atomic configuration in the case of $3$-level atoms interacting with two field modes.

The constants of motion in the Tavis-Cummings model, important to calculate the eigenstates, turn into parity operators when the full Dicke model is considered. As the Hamiltonian in the RWA approximation possesses $\zeta_0$ linear independent constant of motion, $\op{K}_1,\,\op{K}_2,\,\dots,\,\op{K}_{\zeta_0}$, the full Hamiltonian will preserve $\zeta_0$ parity symmetries defined by the operators
\begin{equation}\label{eq.opPi}
\op{\Pi}_\zeta := \exp\left(i\,\pi\,\op{K}_\zeta\right)\,, \quad \zeta=1,\,\dots,\,\zeta_0.
\end{equation}

\begin{table*}
\caption{The operator $\op{K}$ Eq.~(\ref{eq.opK}) as a function of the three free variables for the different $3$-level atomic configurations interacting dipolarly with two modes of radiation, in the RWA approximation. Notice that, for this choice of the free variables, in all cases $\op{K}(0,0,1) = \op{A}_{11}+\op{A}_{22}+\op{A}_{33}$ is the first order Casimir operator, which is the number of atoms in the system, and $\op{K}(1,1,0)$ is the operator of the total number of excitations.}\label{t.opK}
\vspace{2mm}
\begin{tabular}{c | l}
Configuration &  \phantom{\hspace{3.5cm}} Constant of motion\\[1mm] \hline \\[-2mm]
$\Xi$ & \resizebox{0.78\hsize}{!}{$\op{K}(\eta_{1},\eta_{2},\lambda_1) = \eta_{1}\,\op{\nu}_{1}+\eta_{2}\,\op{\nu}_{2} + \lambda_1\,\op{A}_{11} + (\eta_{1}+\lambda_1)\,\op{A}_{22} + (\eta_{1}+\eta_{2}+\lambda_1)\,\op{A}_{33}$}\\[3mm]
$\Lambda$ & \resizebox{0.78\hsize}{!}{$\op{K}(\eta_{1},\eta_{2},\lambda_1) = \eta_{1}\,\op{\nu}_{1}+\eta_{2}\,\op{\nu}_{2} + \lambda_1\,\op{A}_{11} + (\eta_{1}-\eta_{2}+\lambda_1)\,\op{A}_{22} + (\eta_{1}+\lambda_1)\,\op{A}_{33}$}\\[3mm]
$V$ & \resizebox{0.72\hsize}{!}{$\op{K}(\eta_{1},\eta_{2},\lambda_1) = \eta_{1}\,\op{\nu}_{1}+\eta_{2}\,\op{\nu}_{2} + \lambda_1\,\op{A}_{11} + (\eta_{1}+\lambda_1)\,\op{A}_{22} + (\eta_{2}+\lambda_1)\,\op{A}_{33}$}
\end{tabular}
\end{table*}

\subsection{Bases}
\label{bases}

The elements of the basis states are given by a direct product of harmonic oscillators for the field and matter contributions
\begin{equation*}
|\vec{\nu};\vec{b}\ket=|\nu_1,\,\nu_2,\,\dots,\,\nu_\ell\ket\otimes|b_1,\,b_2,\,\dots,\,b_n\ket\,,
\end{equation*} 
with $\nu_s=0,\,1,\,\dots\,,$ and positive integer values $b_k$ which fulfill $b_1+b_2+\cdots+b_n=N_a$, with $\op{A}_{jj} |\vec{\nu};\vec{b}\ket= b_j|\vec{\nu};\vec{b}\ket$.

Exploiting the fact that $\op{K}_\zeta$ are constants of motion in the RWA approximation, and using a set of fixed values $\kappa:=\{k_1,\,k_2,\,\dots,\,k_{\zeta_0}\}$ as indices for the operators $\op{K}_\zeta$, the basis in the RWA approximation may be written as
\begin{equation}\label{eq.basisRWA}
{\cal B}_{\textsc{rwa}}^{(\kappa)} := \left\{|\vec{\nu};\vec{b}\ket \,\Big| \,\op{K}_\zeta|\vec{\nu};\vec{b}\ket=k_\zeta|\vec{\nu};\vec{b}\ket; \,\zeta=1,\,\dots ,\,\zeta_0\right\}\,.
\end{equation}
The dimension of this basis for three-level atoms interacting with two modes of electromagnetic field is given in \ref{s.rwa.dim} for each atomic configuration. We may see that for special values of $k_1$ and $k_2$ the dimension of the basis is equal to the degeneracy of a $3$-dimensional harmonic oscillator with $N_a$ quanta.

The matrix representation of the Hamiltonian~(\ref{eq.H}) in the basis ${\cal B}_{\textsc{rwa}}^{(\kappa)}$ yields the matrix $\op{H}_\textsc{rwa}^{ (\kappa)}$ in the RWA approximation, because the basis~(\ref{eq.basisRWA}) preserves the values of the set $\kappa$, that is, the matrix elements of the counter-rotating terms with respect to the basis ${\cal B}_{\textsc{rwa}}^{(\kappa)}$ vanish. Therefore, a basis for a fixed parity of the full Hamiltonian can be given as the direct sum of the RWA bases, preserving the parity of the values $\kappa$.

Let us denote the parity set of each basis by 
\begin{eqnarray*}\label{eq.parity}
\sigma= \left\{ e^{i \pi k_1},  e^{i \pi k_2}, e^{i \pi k_3},\cdots,  e^{i \pi k_{\zeta_0}}\right\} :={\rm parity}\,(\kappa),
\end{eqnarray*}
where ${\rm parity}\,(\kappa):= \{{\rm parity}\,(k_1),\dots,{\rm parity}\,(k_{\zeta_0})\}$ and ${\rm parity}\,(k)=e\, (o)$ for even (odd) $k$ value. (We have used the label $e$ for {\it even} when the value of the exponential is $+1$, and the label $o$ for {\it odd} when the value of the exponential is $-1$.) So, for a system with $\zeta_0=2$ (i.e., two constants of motion) the basis is divided in four sub-bases  corresponding to the parities $\sigma=ee,\,eo,\,oe$ and $oo$.
Then, if $\kappa_0$ is the parity set of {\it minimum} values which $\kappa$ can take, i.e., $\sigma:={\rm parity}\,(\kappa_0)$,
one can construct a basis with fixed parity for the full matrix Hamiltonian by considering the direct sum of bases of the form ${\cal B}_{\textsc{rwa}}^{(\kappa_0+2j)}$ as follows
\begin{equation*}
{\cal B}_\sigma= \Oplus_{j_1=0}^\infty\,\Oplus_{j_2=0}^\infty\,\cdots\Oplus_{j_{\zeta_0=0}}^\infty\,{\cal B}_{\textsc{rwa}}^{(\kappa_0+2\{j_1,\dots,j_{\zeta_0}\})}\,;
\end{equation*}
the respective matrix Hamiltonian will be denoted by $\op{H}_\sigma$.   

Since the basis of the full Hamiltonian is infinite, it is necessary to truncate the basis to upper values for the set $\kappa_{\rm max}$. A fidelity criterion to do this truncation was proposed recently~\cite{cordero19b}, which allows us to calculate the ground state of the system to a fixed precision. Following this criterion, the number of photons is set in order to have a fidelity between neighbouring states within $10^{-10}$. This, of course, is arbitrary, and may depend on the nature of the problem to be tackled; here we chose this figure as it guarantees an energy difference in neighbouring ground states of less than $10^{-8}$.
%

\section{Quantum Phase Diagram via the Fidelity}
\label{QPD}

Here we discuss the construction of the ground state energy surfaces for both the GTCM and GDM Hamiltonians. For the GTCM Hamiltonian, the energy surfaces are associated to different sets of values for the constants of motion $\kappa=\{ k_1,k_2,\cdots, k_{\zeta_0} \}$ while for the GDM the energy surfaces are associated to the different sets of parities of $\kappa$. One can discuss both cases by labeling the energy surfaces with $\sigma_j$, which will represent fixed values of $\kappa$ for the GTCM while a set of parities of $\kappa$ for the GDM.

In both cases, the ground state energy surface is obtained by comparing the set of energy surfaces with different values of $\sigma_j$ and taking the minimum value at each point in parameter space $(\mu_{ki})$. For each label $\sigma_j$ one has a set of energy surfaces ${\cal E}_{\sigma_j}=\{E_{\sigma_j}^{(0)},\,E_{\sigma_j}^{(1)},\,\dots\}$ as functions of the parameters, with $E_{\sigma_j}^{(r)} \leq E_{\sigma_j}^{(s)}$ for all $r<s$. For each eigenvalue $E_{\sigma_j}^{(r)}$ there is a corresponding eigenstate $|\Psi^{(r)}_{\sigma_j}\ket$ of the Hamiltonian $\op{H}_{\sigma_j}$. Thus, the ground state of the system  $|\Psi_g\ket$ has the energy eigenvalue
\begin{equation}\label{eq.Eg}
E_g = \min\left\{\bigcup_{j}{\cal E}_{\sigma_j}\right\}\,.
\end{equation}
Notice that, in particular for the ground state, this energy value is given by $E_g=\min\{E_{\sigma_1}^{(0)},\,E_{\sigma_2}^{(0)},\,\dots\}$ since $E_{\sigma_j}^{(0)}$ is the ground energy value for each label $\sigma_j$. However one may obtain, in general, the energy surface of excited states by means of 
\begin{equation}
E_r = \min\left\{\bigcup_{j}{\cal E}_{\sigma_j} \Big \backslash \{E_g,E_1,\dots,E_{r-1}\}\right\}\,,
\end{equation} 
where in the set of energy surfaces the lowest $r-1$ energies $\{E_g, E_1, \cdots, E_{r-1} \}$ have been eliminated.  As an example, the first excited state is given by 
\[E_1 = \min\left\{\bigcup_{j}{\cal E}_{\sigma_j} \Big \backslash  \{E_g\}\right\}\,.\]

In this work, we focus our study on the ground state, but we want to stress the fact that a similar analysis can be done for excited states.

\subsection{Finite Quantum Phase Diagram}

From the ground energy surface~(\ref{eq.Eg}), the quantum phase diagram is determined by the set of points where the ground state suffers a sudden change, and this may be detected by means of the calculation of the fidelity concept of quantum information theory, which is defined by 
\begin{equation}\label{eq.fide}
F_\delta(\xi) := |\bra \Psi_g(\xi)|\Psi_g(\xi+\delta)\ket|^2\,;
\end{equation}
this quantity reaches minima at phase transitions because neighbouring states there are dissimilar. 
We have found that the minimum values of the fidelity depend strongly on the selected path. In our case the parameter space has two independent variables $(x_{ij},\,x_{km})$, so we first fix one of them, say $x_{ij}$, and take $\xi$ in eq.~(\ref{eq.fide}) to be $x_{km}$ comparing neighbouring states in that direction; we then vary and compare in the other direction.

In general one finds two types of quantum transitions:  (i) When the fidelity reaches the value $F_\delta(\xi)=0$ it corresponds to a discontinuous transition because the state changes from a Hilbert subspace to an orthogonal one (preserving or not the parity $\sigma$). (ii) When at a minimum $F_\delta(\xi)\neq0$, which corresponds to continuous transitions; in this case the ground state suffers a change, and there is an overlap of the two Hilbert subspaces. 

For the GTCM Hamiltonian with a finite number of particles all the transitions are discontinuous because they correspond to different sets of constants of motion, and are therefore orthogonal. However, in the thermodynamic limit (typically referred to as the limit $N_a \to \infty$) many of these different sets collapse, and the transition emerges as a continuous one (or so called {\it second order transition}) as expected by the variational solution~\cite{cordero13a, cordero13b, nahmad-achar14}.

For a {\it finite} number of particles, we will encounter situations in which the fidelity is close to zero (but not zero), and diminishes as $N_a$ increases, corresponding to the case $F(\xi)=0$ in the limit $N_a\to\infty$ and thus to a discontinuous transition.

On the other hand, there are situations with $F(\xi)\neq 0$ and it either remains different from zero as $N_a$ increases, or reaches zero in the large $N_a$ limit; the former will be {\it stable-continuous transitions}, while the latter {\it unstable-continuous transitions}. In this case, a good measure of the difference between two states, given by density matrices $\rho_A$ and $\rho_B$, across two regions is given by the Bures distance~\cite{bures69, helstrom67}
\begin{equation}
	D_B{}^2(\rho_A,\rho_B) = 2\left( 1 - \sqrt{F(\rho_A,\rho_B)} \right) \,,
\end{equation}
where
\begin{equation}
	F(\rho_A,\rho_B) = \left[ {\rm Tr}\,\sqrt{\sqrt{\rho_A} \rho_B \sqrt{\rho_A}} \right]^2 \,.
\end{equation}

In what follows we apply the fidelity concept to determine the quantum phase diagrams on the corresponding ground state energy surfaces. The study using the Bures distance will be deferred until section~\ref{phase_transitions}.

\subsection{$3$-Level Atoms}
\label{3-level}

We consider the case of a $3$-level system ($n=3$) interacting with two modes ($\ell=2$) of radiation. For this case there are three atomic configurations available, called $\Xi$, $\Lambda$, and $V$ after their schematic structures. The Hamiltonian (with $\hbar=1$ and $\omega_1=0$) and the two symmetries for each atomic configuration are:\\[1mm]

{\bf $\Xi$-configuration.-}\\
\begin{eqnarray}
\op{H}_\Xi &=&  \Omega_{1}\, \op{\nu}_{1}+\Omega_{2}\, \op{\nu}_{2}+ \omega_2 \, \op{A}_{22}+ \omega_3 \, \op{A}_{33}\nonumber \\[3mm]
&-& \frac{\bar{\mu}_{12}^{(1)}}{\sqrt{N_a}}\,x_{12}^{(1)}\, \left(\op{A}_{12}+\op{A}_{21}\right)\left(\op{a}_{1} + \op{a}_{1}^\dag\right)\nonumber\\[3mm]
&-& \frac{\bar{\mu}_{23}^{(2)}}{\sqrt{N_a}}\,x_{23}^{(2)}\, \left(\op{A}_{23}+\op{A}_{32}\right)\left(\op{a}_{2} + \op{a}_{2}^\dag\right)\,,
\end{eqnarray}
and from table~\ref{t.opK} one has two independent symmetries $\op{\Pi}_j = \exp(i\,\pi \op{K}_j)$, which we choose to be
\numparts
\label{eq.k1k2X}
\begin{itemize}
	\item[]
\begin{eqnarray}
\op{K}_1 &=& \op{\nu}_1 + \op{\nu}_2 + \op{A}_{22} + 2\, \op{A}_{33}\,,
\end{eqnarray}
	\item[]
\begin{eqnarray}
\op{K}_2 &=& \op{\nu}_2 + \op{A}_{33}\,.
\end{eqnarray}
\end{itemize}
\endnumparts
$\op{K}_1$ stands for the total number of excitations in the system, and has integer eigenvalues $k_1=0,\,1,\,\dots$, while $\op{K}_2$ plays the role of the excitations related to a $2$-level  subsystem formed by the second and third atomic levels with the mode $\Omega_2$; for a fixed value $k_1$ the operator $\op{K}_2$ may take values $k_2=0,\,1,\,\dots,\,k_1$.\\

{\bf $\Lambda$-configuration.-}\\
\begin{eqnarray}
\op{H}_\Lambda &=&  \Omega_{1}\, \op{\nu}_{1}+\Omega_{2}\, \op{\nu}_{2}+ \omega_2 \, \op{A}_{22}+ \omega_3 \, \op{A}_{33}\nonumber \\[3mm]
&-& \frac{\bar{\mu}_{13}^{(1)}}{\sqrt{N_a}} \,x_{13}^{(1)}\,\left(\op{A}_{13}+\op{A}_{31}\right)\left(\op{a}_{1} + \op{a}_{1}^\dag\right)\nonumber\\[3mm]
&-& \frac{\bar{\mu}_{23}^{(2)}}{\sqrt{N_a}}\,x_{23}^{(2)}\, \left(\op{A}_{23}+\op{A}_{32}\right)\left(\op{a}_{2} + \op{a}_{2}^\dag\right)\,,
\end{eqnarray}
and we chose the symmetry operators 
\numparts
\label{eq.k1k2L}
\begin{itemize}
	\item[]
\begin{eqnarray}
\op{K}_1 &=& \op{\nu}_1 + \op{\nu}_2  +  \op{A}_{33}\,,
\end{eqnarray}
	\item[]
\begin{eqnarray}
\op{K}_2 &=& \op{\nu}_2 + \op{A}_{11} + \op{A}_{33} \,.
\end{eqnarray}
\end{itemize}
\endnumparts
Here $\op{K}_1$ defines the total number of excitations, with integer eigenvalues $k_1=0,\,1,\,\dots$, and for a fixed value of $k_1$ the operator $\op{K}_2$ can take values $k_2=0,\,1,\,\dots,\,k_1+N_a$.\\[1mm]

{\bf $V$-configuration.-}\\
\begin{eqnarray}
\op{H}_V &=&  \Omega_{1}\, \op{\nu}_{1}+\Omega_{2}\, \op{\nu}_{2}+ \omega_2 \, \op{A}_{22}+ \omega_3 \, \op{A}_{33}\nonumber \\[3mm]
&-& \frac{\bar{\mu}_{12}^{(1)}}{\sqrt{N_a}}\,x_{12}^{(1)}\, \left(\op{A}_{12}+\op{A}_{21}\right)\left(\op{a}_{1} + \op{a}_{1}^\dag\right)\nonumber\\[3mm]
&-& \frac{\bar{\mu}_{13}^{(2)}}{\sqrt{N_a}}\,x_{13}^{(2)}\, \left(\op{A}_{13}+\op{A}_{31}\right)\left(\op{a}_{2} + \op{a}_{2}^\dag\right)\,,
\end{eqnarray}
and we chose the independent symmetry operators 
\numparts
\label{eq.k1k2V}
\begin{itemize}
	\item[]
\begin{eqnarray}
\op{K}_1 &=& \op{\nu}_1 + \op{A}_{22}\,,
\end{eqnarray}
	\item[]
\begin{eqnarray}
\op{K}_2 &=& \op{\nu}_2  + \op{A}_{33} \,.
\end{eqnarray}
\end{itemize}
\endnumparts
$\op{K}_1$ denotes the number of excitations of a $2$-level subsystem formed by the first and second atomic levels interacting with radiation mode $\Omega_1$, which takes integer eigenvalues $k_1=0,\,1,\,\dots$, and $\op{K}_2$ is the number of excitations of the other $2$-level subsystem formed by the first and third atomic levels interacting with mode $\Omega_2$, with values $k_2=0,\,1\,\dots$. In this case $\op{K}_1+\op{K}_2$ corresponds to the total number of excitations of the system.

The relationship between the values $k_1$ and $k_2$ is useful in order to build the cutoff basis ${\cal B}_{\sigma}$ of the full Hamiltonian~(\ref{eq.H}).
%
\begin{figure}
\begin{center}
\includegraphics[width=0.7\linewidth]{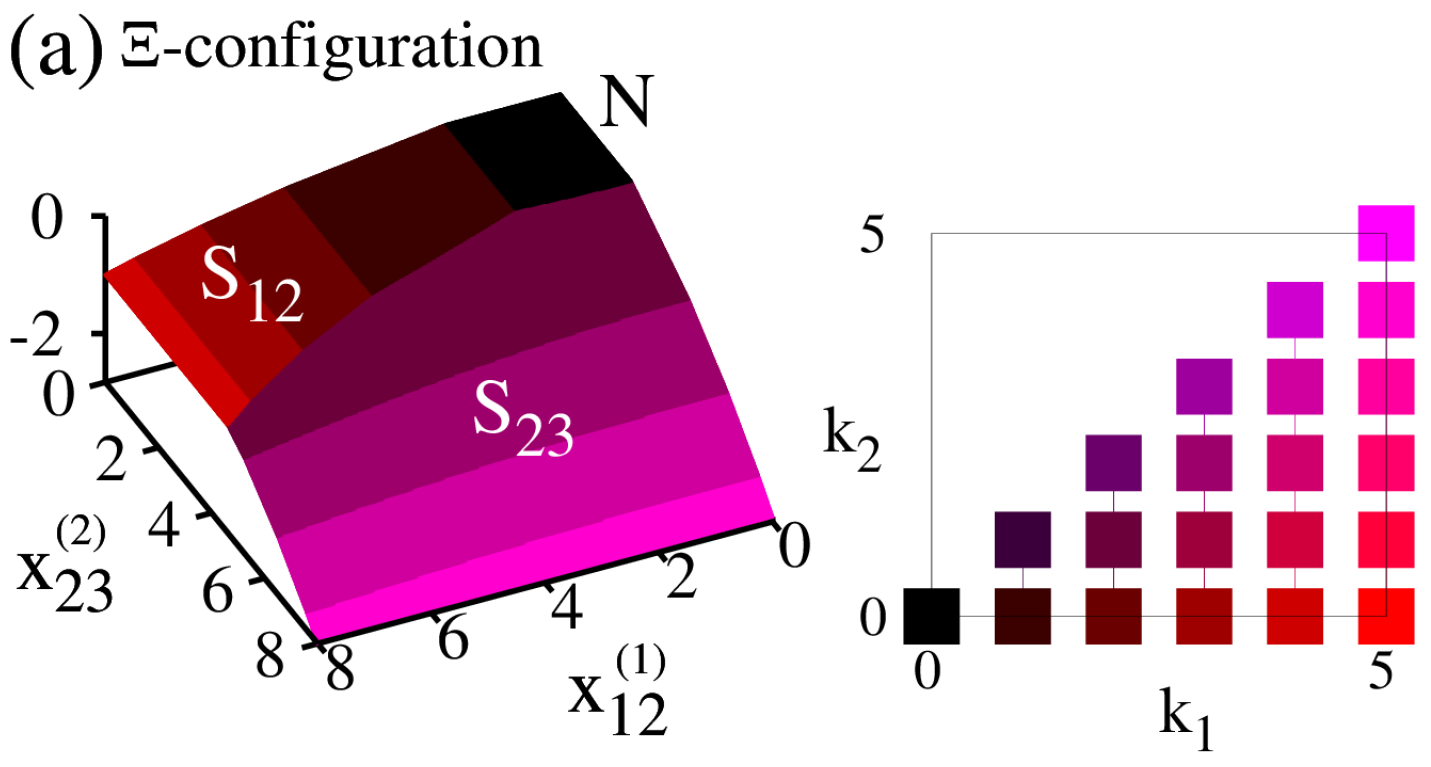}\\[5mm]
\includegraphics[width=0.7\linewidth]{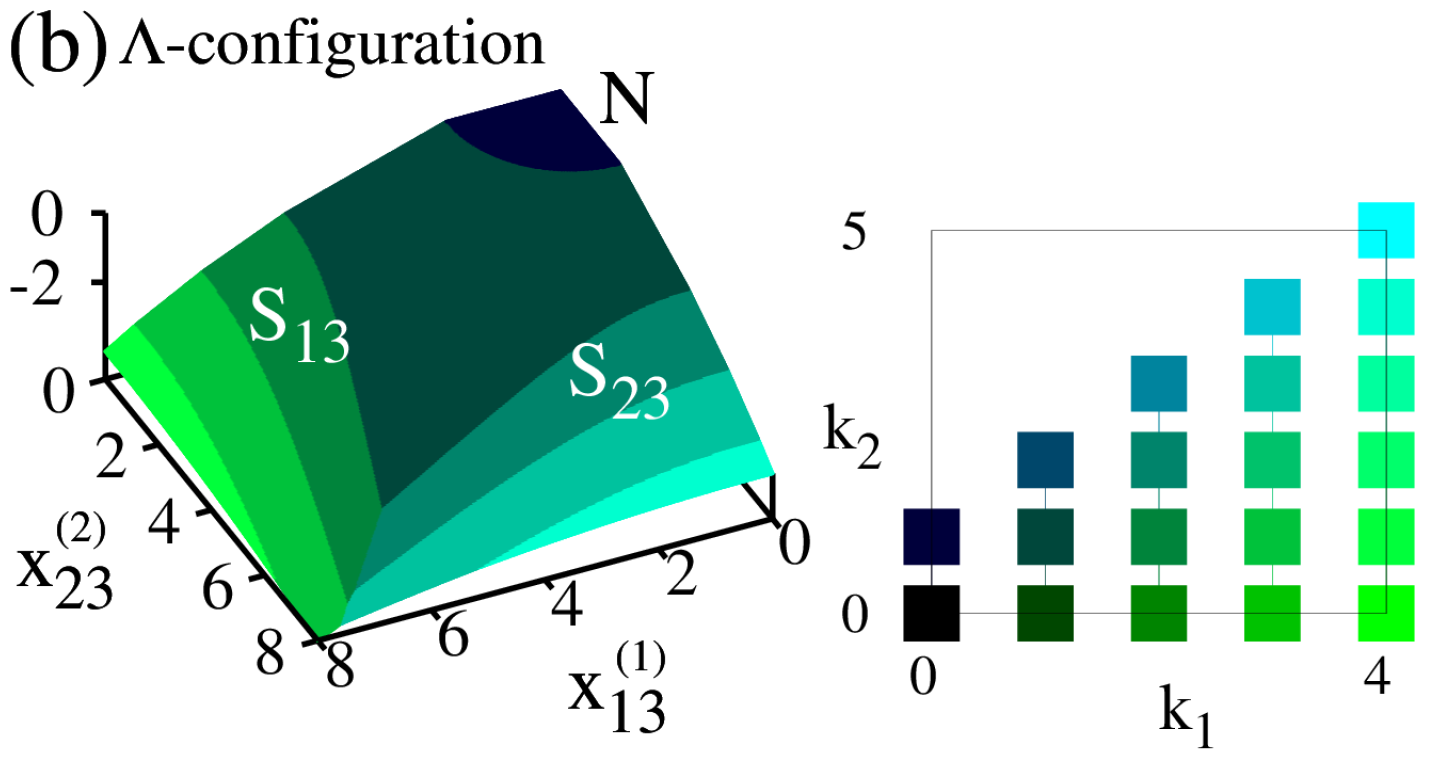}\\[5mm]
\includegraphics[width=0.7\linewidth]{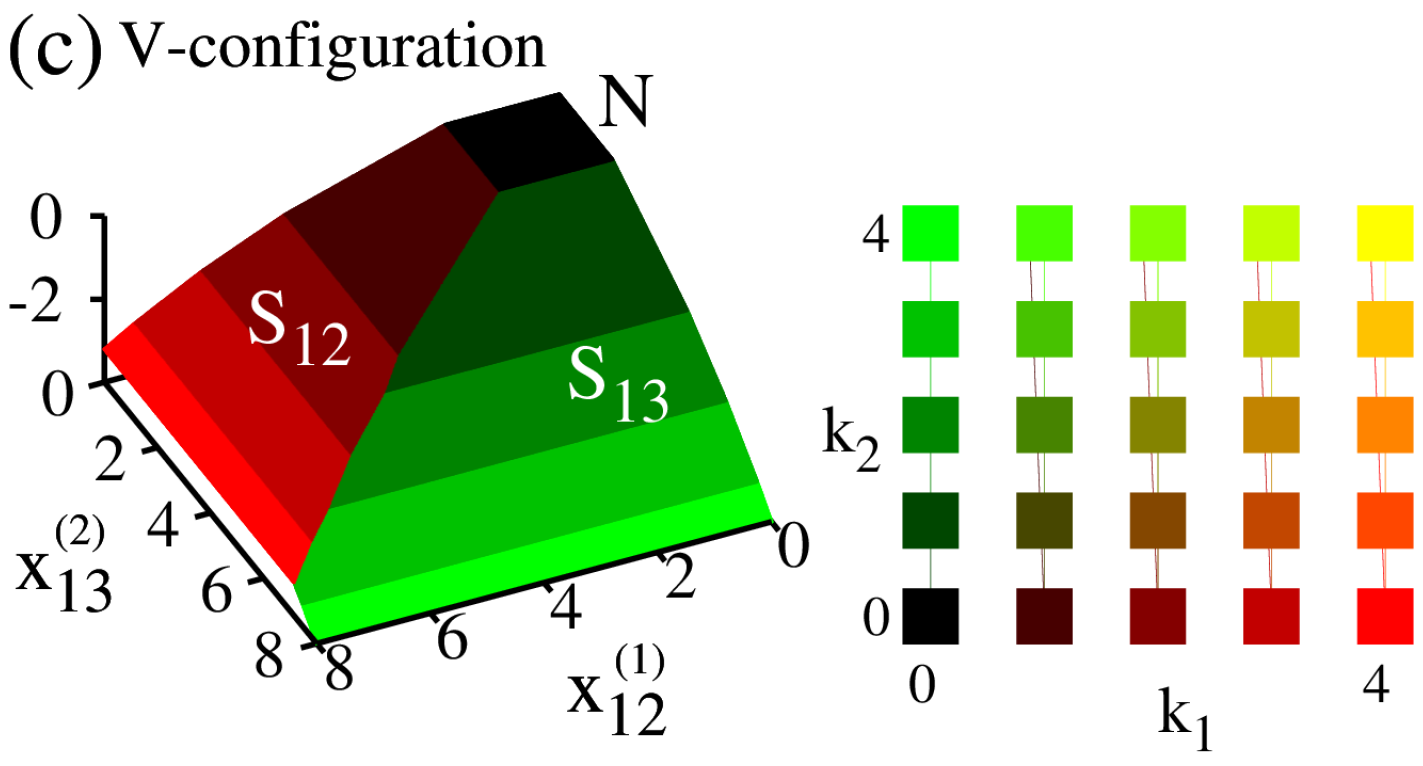}
\end{center}
\caption{(Colour online.) The ground state energy surface per particle as a function of the control parameters $x_{ij}$ is shown, for the GTCM model and $N_a=1$. The values of $k_1$ and $k_2$ are given in the coloured legends. (a) For the $\Xi$-configuration the parameters are:  $\Omega_1=0.25, \, \Omega_2=0.75,\, \omega_1=0,\, \omega_2=0.25$ and $\omega_3=1$. (b) For the $\Lambda$-configuration,  $\Omega_1=1, \, \Omega_2=0.9,\, \omega_1=0,\, \omega_2=0.1$ and $\omega_3=1$. (c) For the $V$-configuration, $\Omega_1=0.8, \, \Omega_2=1,\, \omega_1=0,\, \omega_2=0.8$ and $\omega_3=1$. In all cases the energy and frequencies are in units of $\omega_3$.}
\label{f.eminRWANa1}  
\end{figure}
%
\begin{figure}
\begin{center}
\includegraphics[width=0.7\linewidth]{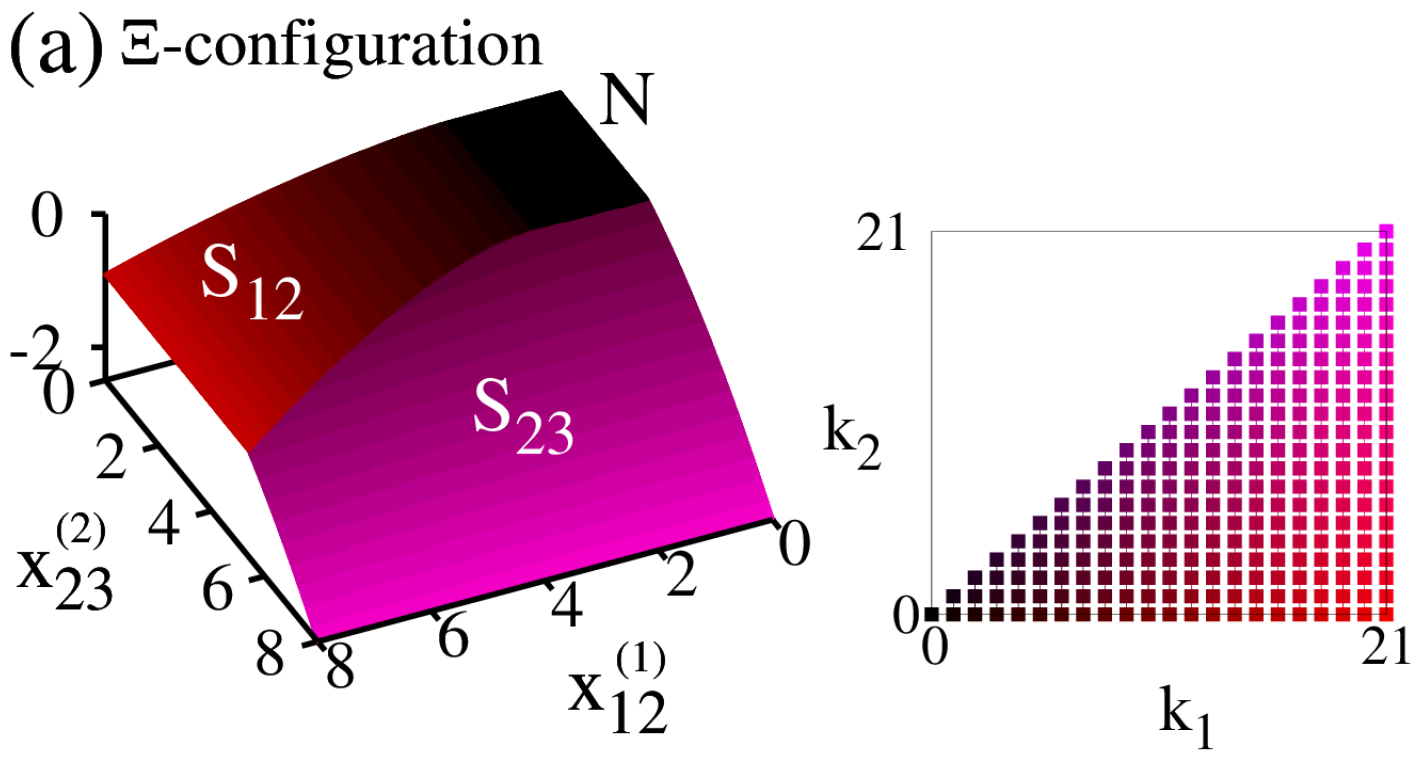}\\[5mm]
\includegraphics[width=0.7\linewidth]{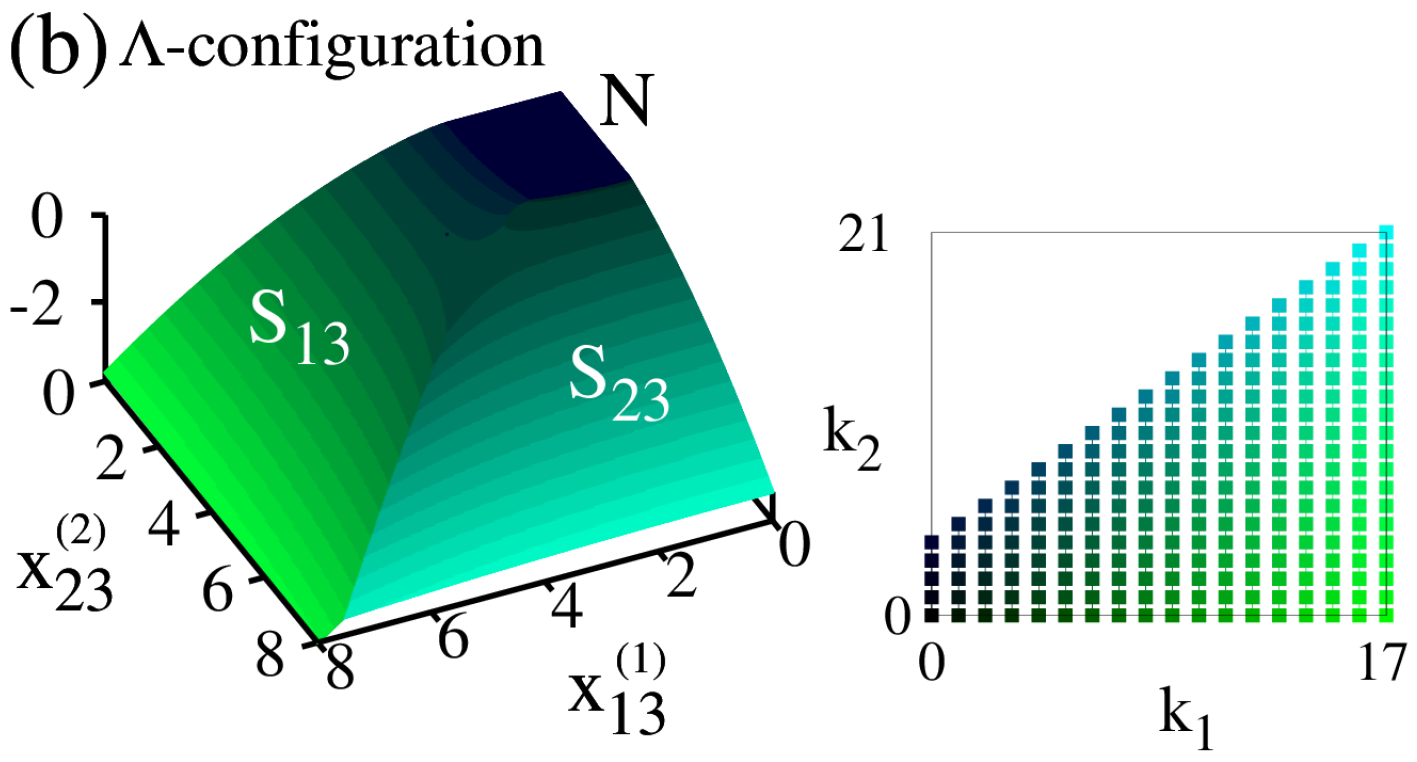}\\[5mm]
\includegraphics[width=0.7\linewidth]{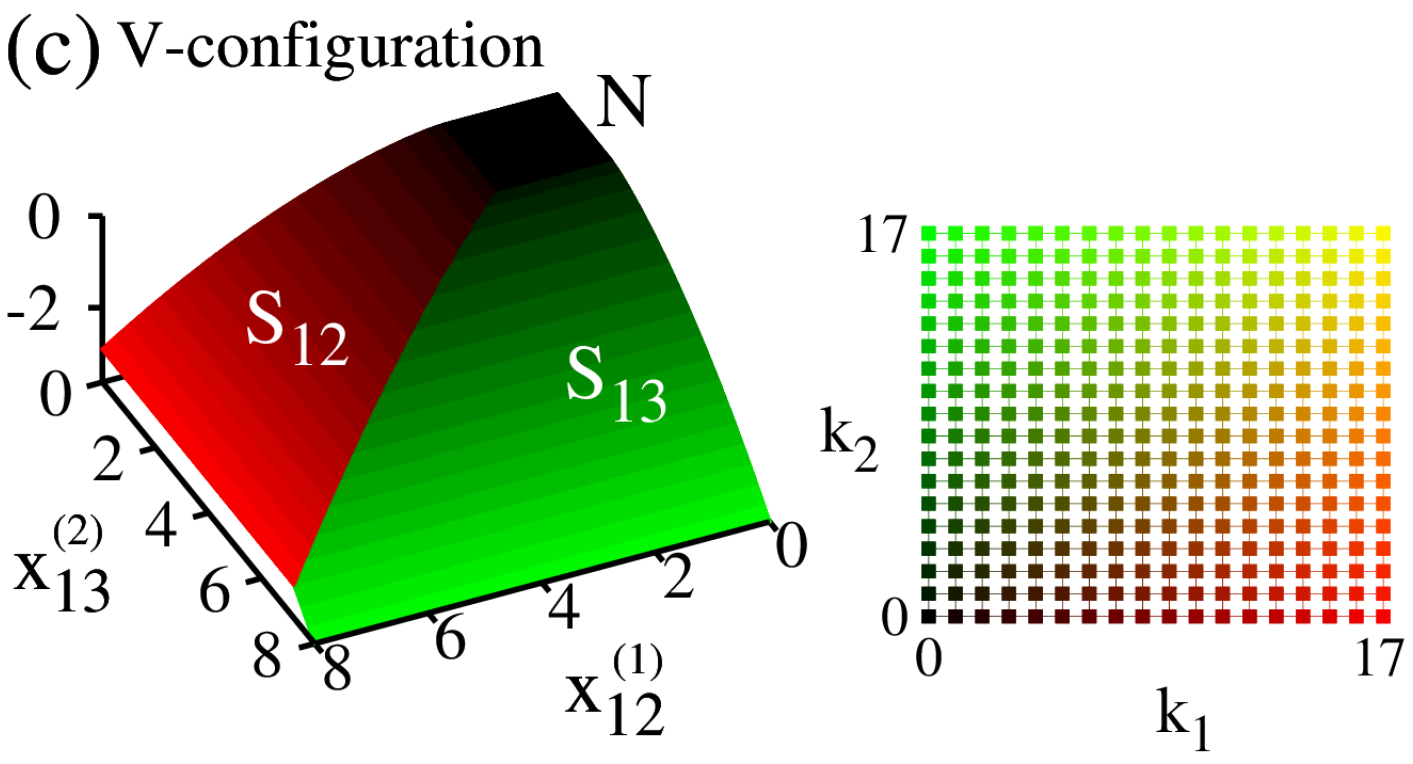}
\end{center}
\caption{(Colour online.) The ground state energy surface per particle as a function of the control parameters $x_{ij}$ is displayed, for the GTCM model and $N_a=4$. The values of $k_1$ and $k_2$ are given in the coloured legends. The configurations are (a)  $\Xi$, (b) $\Lambda$ and (c) $V$. Parameters are the same as in Fig.~\ref{f.eminRWANa1}.}
\label{f.eminRWANa4}  
\end{figure}

\subsection{Generalised Tavis-Cummings Model}
\label{GTCM}

In the generalised Tavis-Cummings model the corresponding operators $\op{K}_1$ and $\op{K}_2$ are constants of motion. For fixed values of the dimensionless coupling matter-field parameters $x_1$ and $x_2$ one needs to find the eigenvalues of $\op{K}_1$ and $\op{K}_2$ for which the energy surfaces reach their minimum value. 

We calculate all the energy surfaces ${\cal E}_\kappa$ with values 
\[
\kappa=\{k_1,\,k_2\} \, ,
\]
by considering the integer intervals $0\leq k_1 \leq k_{1{\rm max}}$ and $0\leq k_2 \leq k_{2{\rm max}}$, where the values $\{k_{1{\rm max}},\,k_{2{\rm max}}\}$ are those required to describe the ground state energy surface for the maximum values of the dimensionless coupling parameters $x_1$ and $x_2$ in the region of study of the phase space of the system. The maximum values of these $k_j$ may be calculated from two-level systems as was described previously~\cite{cordero19b}. Thus, to obtain the ground energy surface involves the evaluation of a maximum of $(k_{1{\rm max}}+1)(k_{2{\rm max}}+1)$ surfaces in the region $[0,x_1]\times[0,x_2]$.

In this work we found, after analyzing the corresponding set of energy surfaces ${\cal E}_\kappa$ for each atomic configuration ($\Xi, \Lambda$ and $V$), that the pair of values $\{k_1,k_2\}$ involved in the ground state energy surface obeys a simple relationship which is obtained from those values $\{k_1,k_2\}$ arising from the two-level subsystem. With these constraints we then determine the eigenvalues $\{k_1,k_2\}$ that the constants of motion $\op{K}_1$ and $\op{K}_2$ can take.

As an example consider the $\Xi$-configuration. The constants of motion are $\op{K}_1=\op{\nu}_1+\op{\nu}_2+\op{A}_{22}+2\op{A}_{33}$ and $\op{K}_2=\op{\nu}_2+\op{A}_{33}$ as given in Eq.~(\ref{eq.k1k2X}), with eigenvalues $k_1$ and $k_2$ respectively. This configuration has two subsystems, namely, ${\cal H}_{12}$ for the mode $\Omega_1$ and ${\cal H}_{23}$ for the mode $\Omega_2$. Therefore, if one considers only the contribution of the subsystem ${\cal H}_{12}$ we fix the eigenvalues $\nu_2=0$ and $n_3=0$ of the operators $\op{\nu}_2$ and $\op{A}_{33}$, having for this reduction the eigenvalues $\{k_1,k_2\}=\{k_1,0\}$ with $k_1=0,1,\dots$. On the other hand, for the subsystem ${\cal H}_{23}$ we fix the eigenvalues $\nu_1=0$ and $n_1=0$ of the operators $\op{\nu}_1$ and $\op{A}_{11}$, and then we find the eigenvalues $\{k_1,k_2\}=\{k_2+N_a,k_2\}$ with $k_2=0,1,\dots$; in this last expression we used the fact that for the subsystem ${\cal H}_{23}$ the relationship $\op{A}_{22}+\op{A}_{33}=N_a$ must be fulfilled. Thus, only the energy surfaces ${\cal E}_\kappa$ with labels $\kappa$ of the form $\{k_1,0\}$ and  $\{k_2+N_a,k_2\}$ contribute to the ground energy surface, simplifying the calculation.

Using the same rule we find the labels $\kappa$ for the surfaces ${\cal E}_\kappa$ which contribute to the ground state energy surface in each atomic configuration, these are:

\begin{itemize}
\item $\kappa=\{k_1,\,0\},\, \{k_2+N_a,\,k_2\}$ for the $\Xi$-configuration,
\item $\kappa=\{k_1,\,N_a\},\, \{k_2,\,k_2\}$ for the $\Lambda$-configuration,
\item $\kappa=\{k_1,\,0\},\, \{0,\,k_2\}$ for the $V$-configuration,
\end{itemize}
where $k_1=0,\,1,\,\dots,\,k_{1{\rm max}}$ and $k_{2}=0,\,1,\,\dots,\,k_{2{\rm max}}$ for the region $[0,x_1]\times[0,x_2]$. The procedure just described resembles that to obtain the reduced bases in the GDM~\cite{cordero19}.

In figures~\ref{f.eminRWANa1}  and \ref{f.eminRWANa4} the energy surface per particle is shown as a function of the dimensionless dipolar strengths, for each atomic configuration: $\Xi$, Figs.~\ref{f.eminRWANa1}(a) and~\ref{f.eminRWANa4}(a); $\Lambda$, Figs.~\ref{f.eminRWANa1}(b) and~\ref{f.eminRWANa4}(b); and $V$, Figs.~\ref{f.eminRWANa1}(c) and~\ref{f.eminRWANa4}(c). Each surface is coloured according to the coloured values of $\{k_1,\,k_2\}$ shown at the right of each plot. Thus the separatrices of the energy surface take place at the loci where a sudden change of colour happens, and at these loci the ground state changes from one Hilbert subspace to another. Since $k_1$ represents the total number of excitations in the $\Xi$ and $\Lambda$ configurations, their normal region is characterised by $k_1=0$. Similarly, for the $V$ configuration the normal region is given by $k_1+k_2=0$.

For the $\Xi$-configuration in Fig.~\ref{f.eminRWANa1}(a) the normal region was determined by the values $\{k_1,k_2\}=\{0,0\}$ (black region) where the cavity is in the vacuum state and the atom is in its lower atomic level. The collective region is divided into two collective sub-regions: the $S_{12}$ sub-region where the mode $\Omega_1$ dominates, is characterised by the set of values $\{k_1,0\}$ and yields values for the basis states with $\nu_2=0$ and $b_3=0$, hence there are only transitions between the atomic levels $\omega_1\rightleftharpoons \omega_2$; and the $S_{23}$ sub-region with dominant mode $\Omega_2$, characterised by the set of values $\{k_2+N_a,k_2\}$ and transitions $\omega_2\rightleftharpoons \omega_3$. Notice that states with non-zero contribution of the mode $\Omega_1$ should satisfy $N_a=\nu_1+b_2+b_3=1$. 

For the $\Lambda$-configuration in Fig.~\ref{f.eminRWANa1}(b) the normal region was determined by the values $\{k_1,k_2\}=\{0,1\}$ (dark blue region) while the collective region is divided into two sub-regions, namely $S_{13}$ and $S_{23}$ where the modes $\Omega_1$ and $\Omega_2$ dominate the atomic transitions respectively. Here, the sub-region $S_{13}$ is characterised by $\{k_1,1\}$ (the states with non-zero contribution of the mode $\Omega_2$ satisfy $N_a=\nu_2+b_1+b_3=1$), and the sub-region $S_{23}$ by the values $\{k_2,k_2\}$ (the states with non-zero contribution of the mode $\Omega_1$ fulfill $\nu_1-b_1=0$). 

\begin{figure}
\begin{center}
\includegraphics[width=0.6\linewidth]{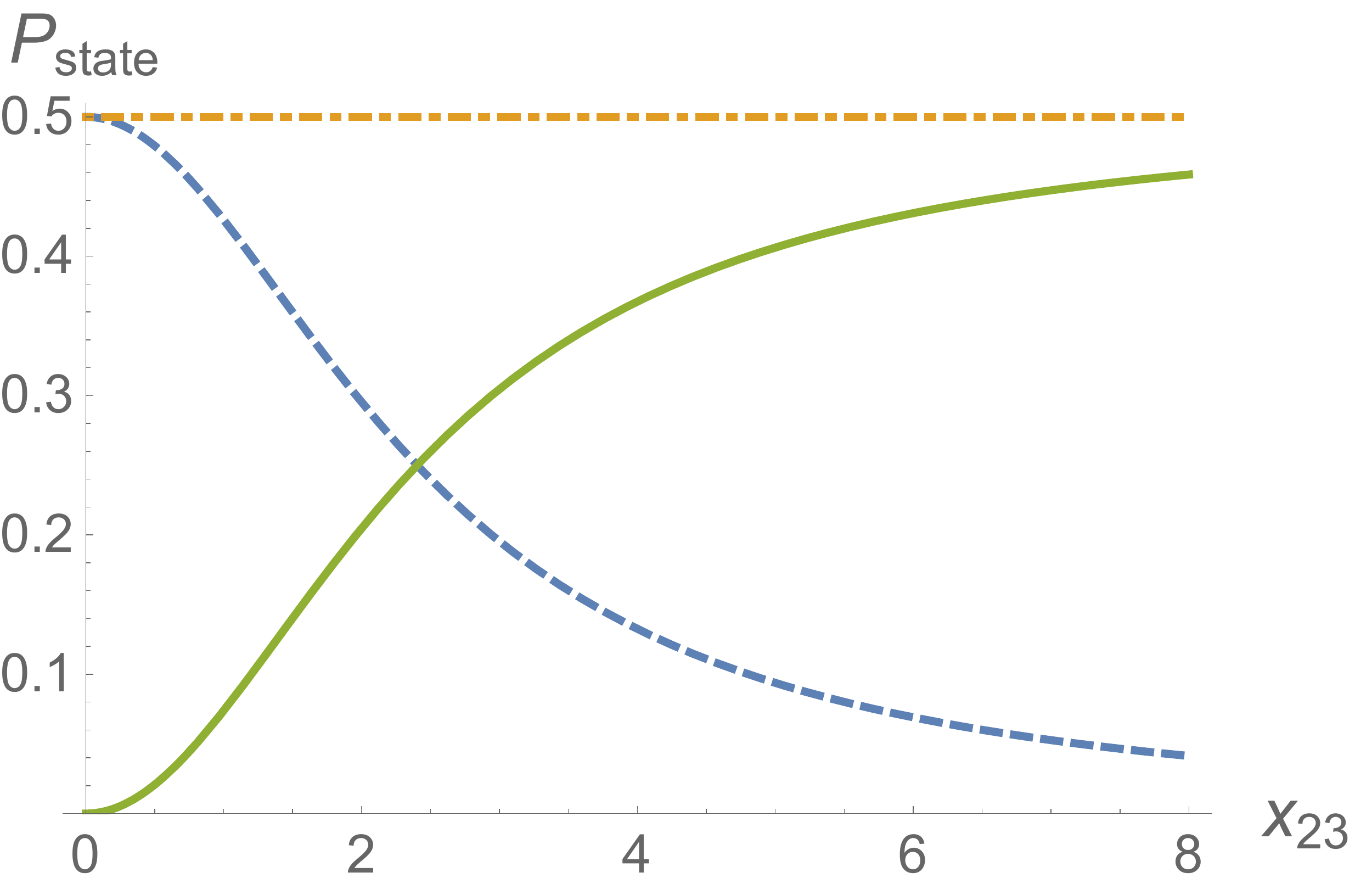}
\end{center}
\caption{(Colour online.) Probability that the ground state is $\vert 1,0;1,0,0\rangle$ (dashed blue line), $\vert 0,1;0,1,0\rangle$ (continuous green line), or $\vert 0,0;0,0,1\rangle$ (dot-dashed orange line), as a function of the dipolar strengths for the atomic configuration $\Lambda$ and parameter values are given in figure~\ref{f.eminRWANa1}, for $x_{13}=3$.}
\label{casoLk1eqk2}  
\end{figure}

The special case $\{k_1,k_2\}=\{1,1\}$ [dark cyan in the colour legend of figure~\ref{f.eminRWANa1}(b)] corresponds to linear combinations of the states $\vert 1,0;1,0,0\rangle$, $\vert 0,0;0,0,1\rangle$, and $\vert 0,1;0,1,0\rangle$, for all the values shown of $x_{13}$ and $x_{23}$, where the notation $|\nu_1,\nu_2;b_1,b_2,b_3\rangle$ has been used. These linear combinatios are of harmonic oscillator states, which is the basis used for diagonalisation. Superpositions of the first two states lie in the region $S_{13}$, and those of the last two states in the region $S_{23}$. The state $\vert 0,0;0,0,1\rangle$ lies in both regions, and has no contributions of any kind of photons. These regions are not shown divided by a separatrix since the figure is drawn according to values of $\{k_1,k_2\}=\{1,1\}$. The transition is made explicit when we draw the probability that the ground state is in one of the first or third states above, as in figure~\ref{casoLk1eqk2}. In this last figure, the dashed (blue) line gives the probability for $\vert 1,0;1,0,0\rangle$ to be the ground state, the continuous (green) line that of $\vert 0,1;0,1,0\rangle$, and the dot-dashed (orange) line that of the state $\vert 0,0;0,0,1\rangle$. We see that in different regions we have the superpositions mentioned above with higher probabilities.
%

For the $V$-configuration Fig.~\ref{f.eminRWANa1}(c), the normal region is that determined by the values $\{k_1,k_2\}=\{0,0\}$ (black region) while the collective regions are $S_{12}$ with values for $\{k_1,0\}$ and $S_{13}$ with values for $\{0,k_2\}$. This figure exhibits clearly the polychromatic behaviour obtained in the variational solution~\cite{cordero15}.  

Similar results are obtained when the number of particles increases as shown in figure~\ref{f.eminRWANa4} for the three atomic configurations with $N_a=4$ particles. Notice that the shape of the normal region for the $\Lambda$-configuration becomes more rectangular when the number of particles grows, as expected due to the variational solution. 
%
\begin{figure}
\begin{center}
\includegraphics[width=0.55\linewidth]{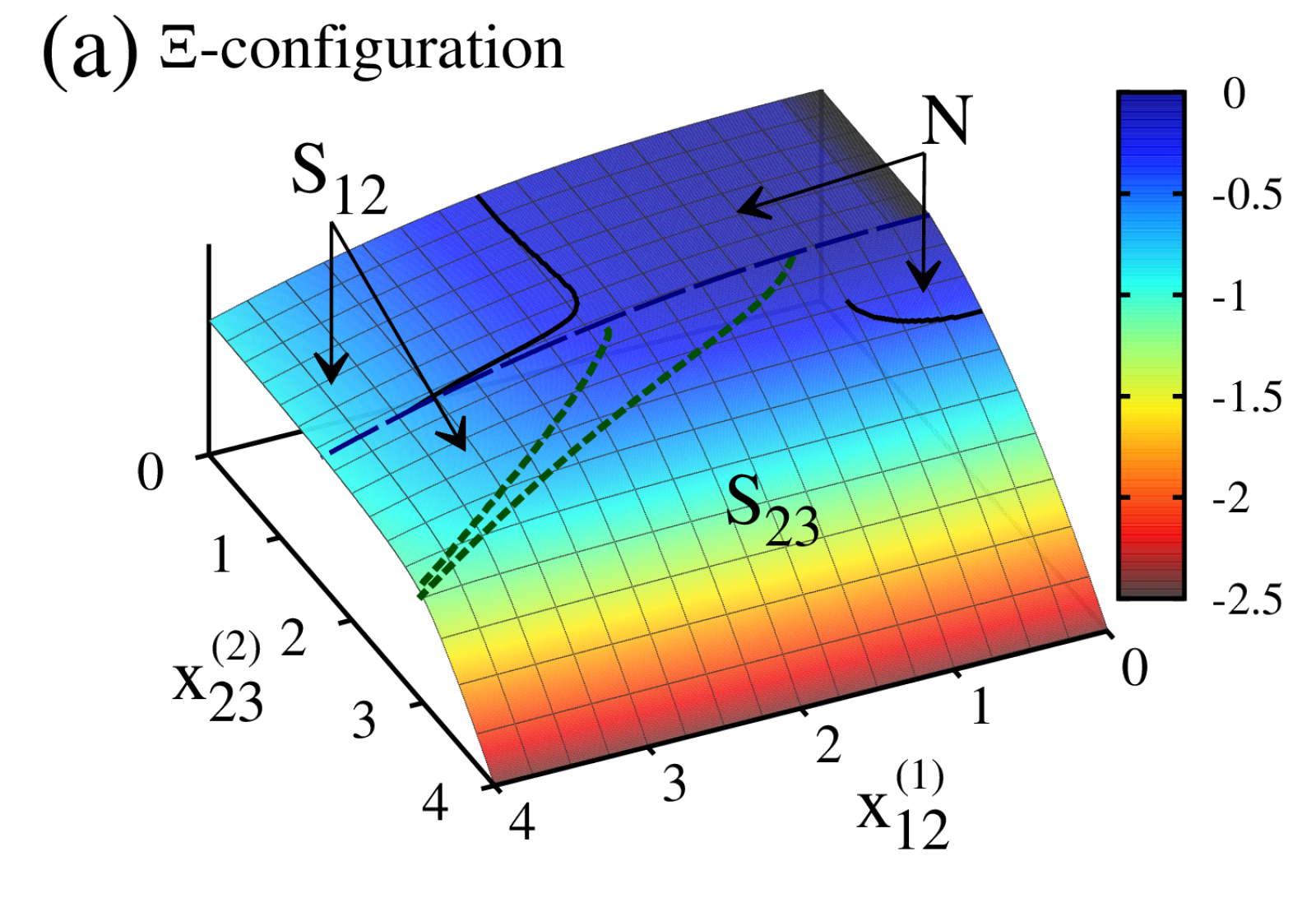}\\
\includegraphics[width=0.55\linewidth]{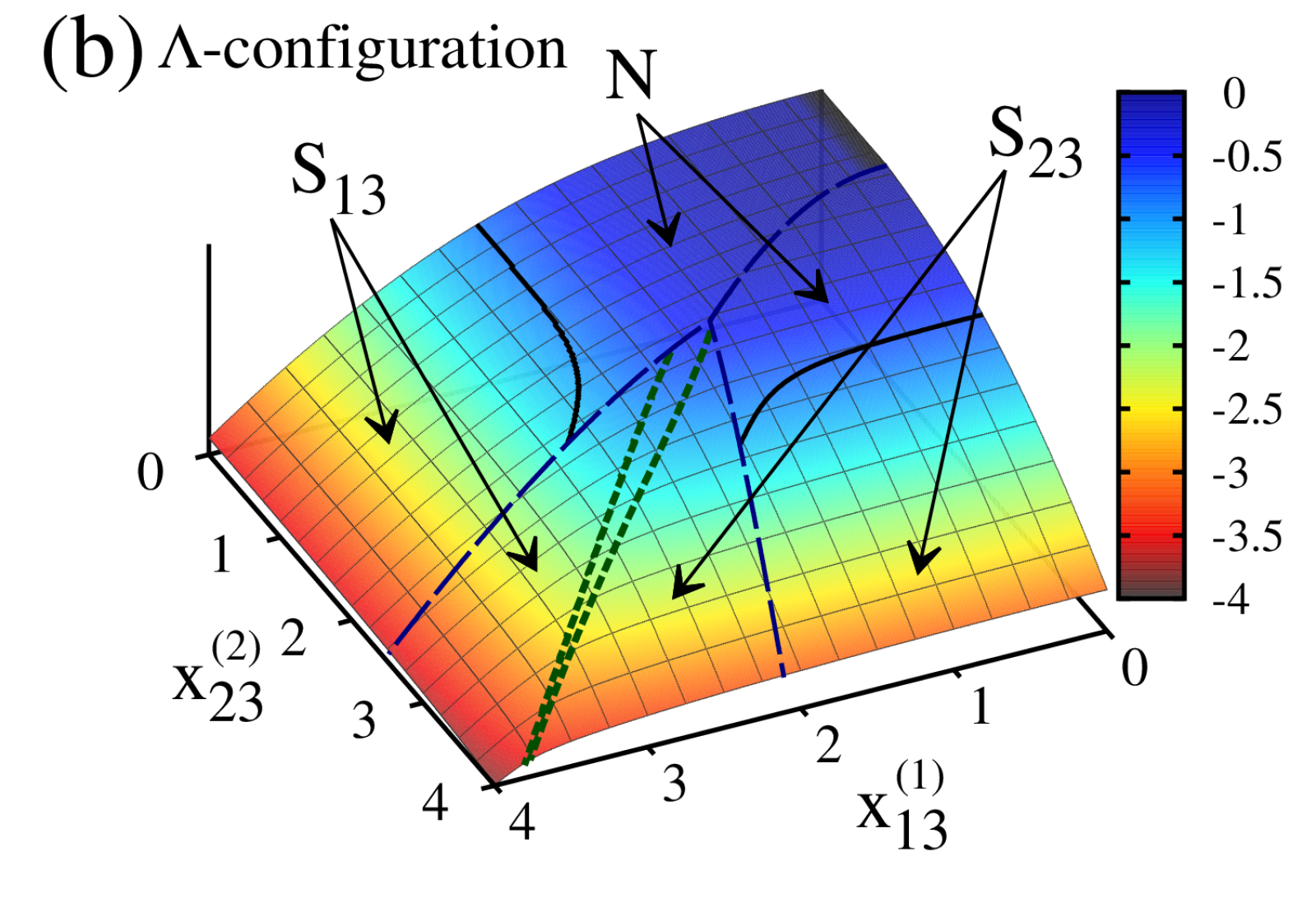}\\
\includegraphics[width=0.55\linewidth]{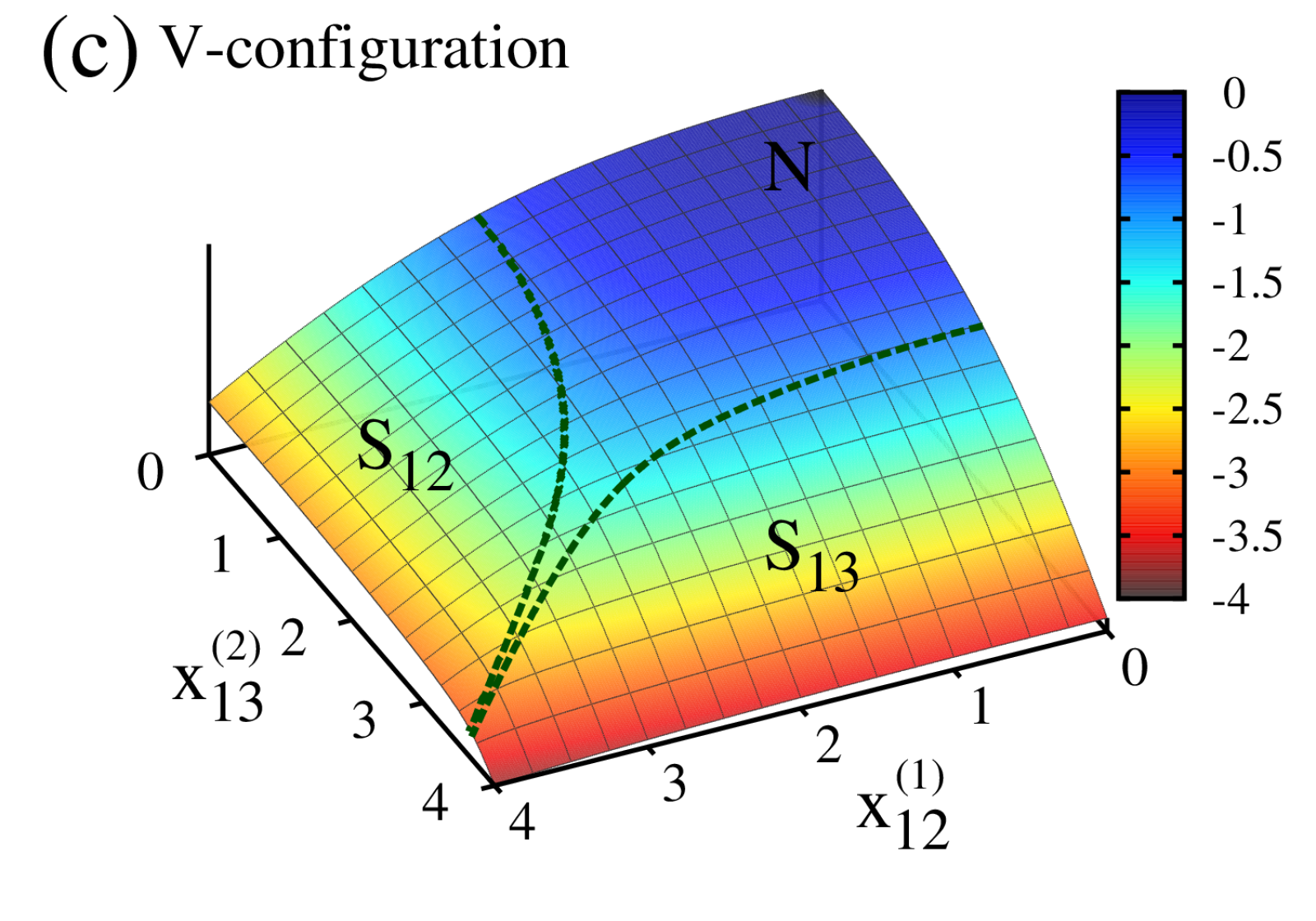}
\end{center}
\caption{(Colour online.) Ground state energy surface and quantum separatrices (lines) for a single particle is shown for $3$-level atoms as function of the dimensionless dipolar strengths. The atomic configurations are: (a) $\Xi$, (b) $\Lambda$ and (c) $V$. There are $3$ types of lines in the figure:
across continuous lines there is no change of parity; across dashed lines there is a change of parity, these appear for $N_a$ odd, and transitions across these lines are discontinuous; across dotted lines there is no change of parity, the transition is continuous (of second order) while the two lines are apart, and become first-order transitions when they coalesce. Parameter values are given in figure~\ref{f.eminRWANa1}.}
\label{f.eminNa1}  
\end{figure}
%
\begin{figure}
\begin{center}
\includegraphics[width=0.55\linewidth]{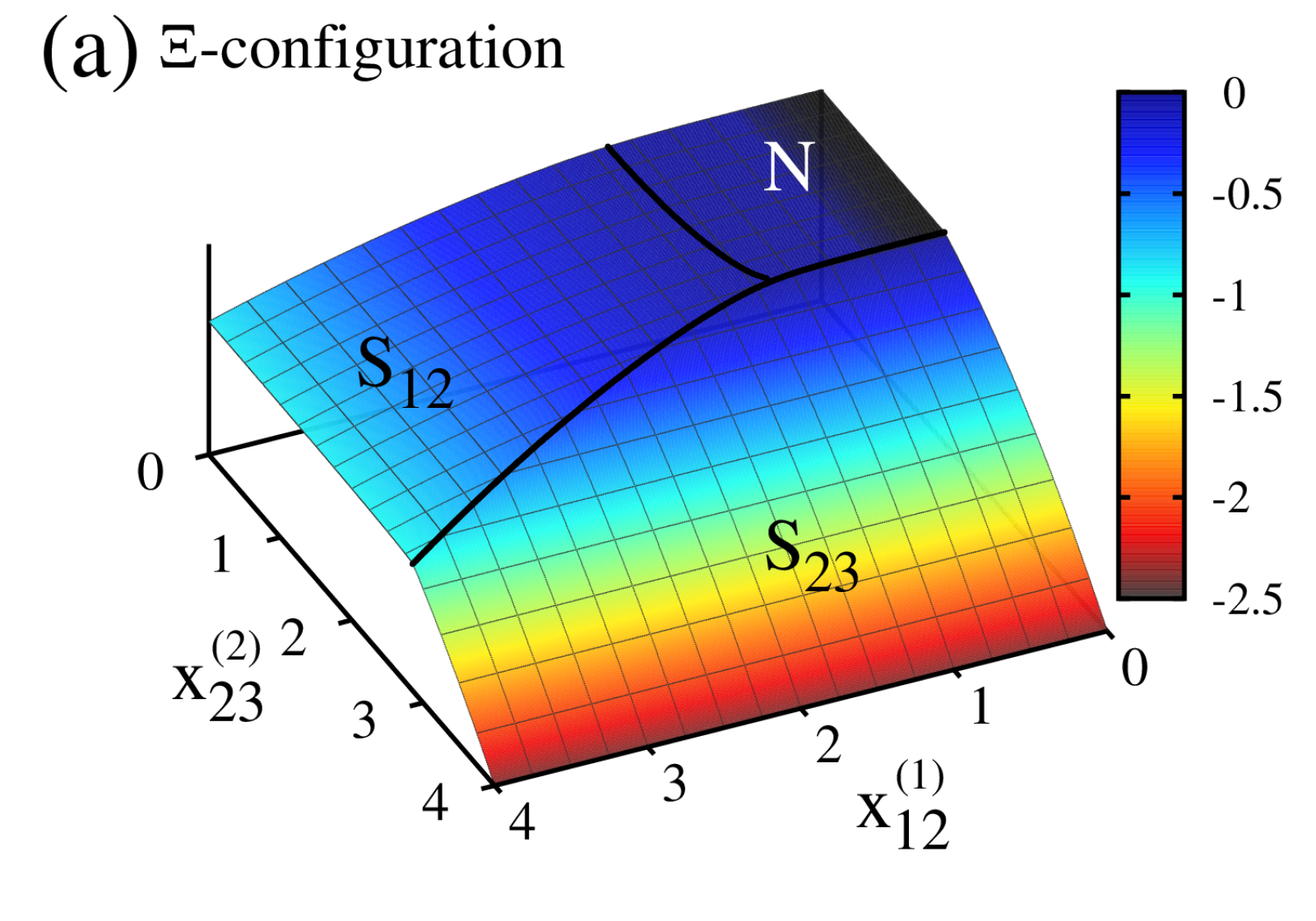}\\
\includegraphics[width=0.55\linewidth]{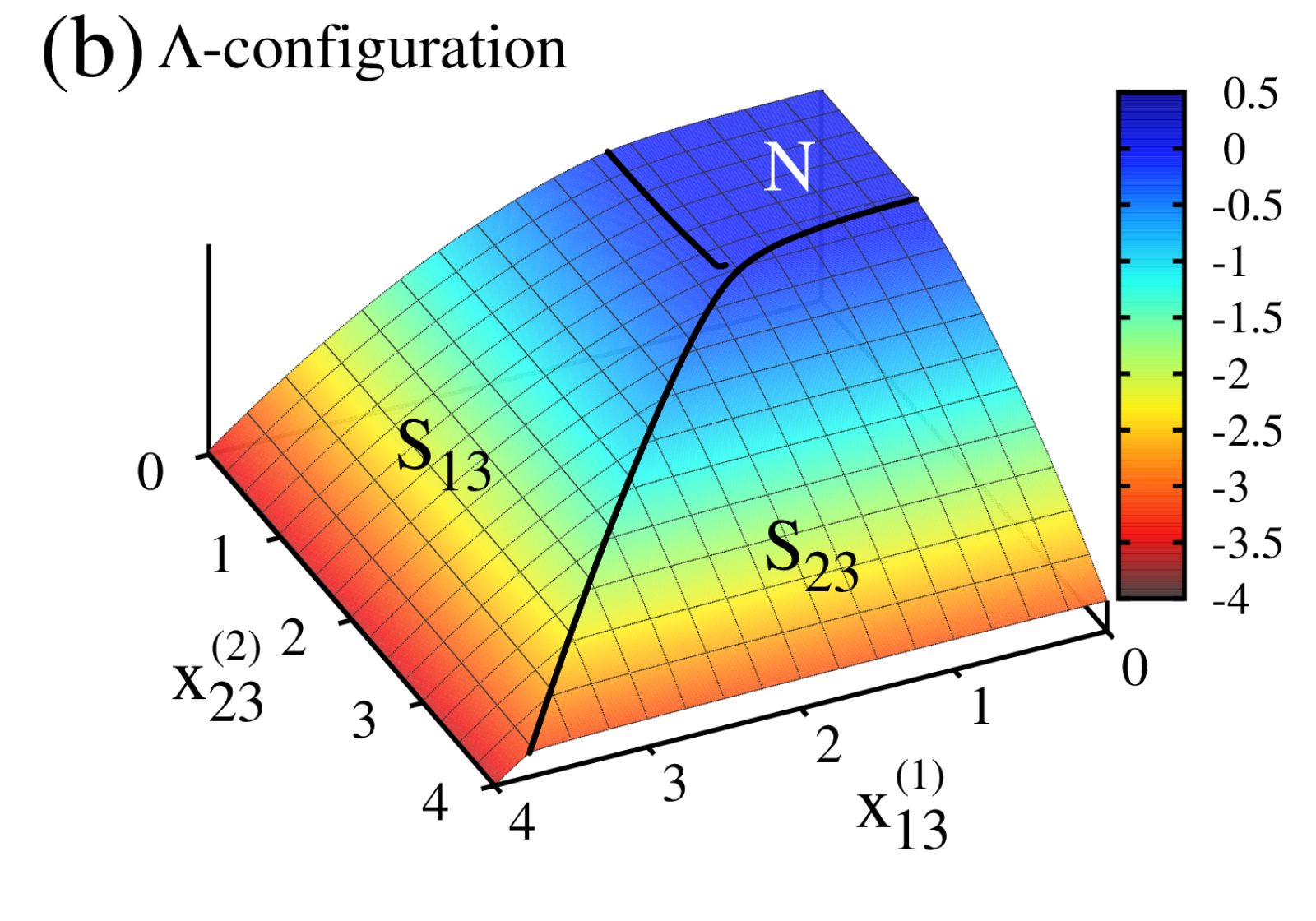}\\
\includegraphics[width=0.55\linewidth]{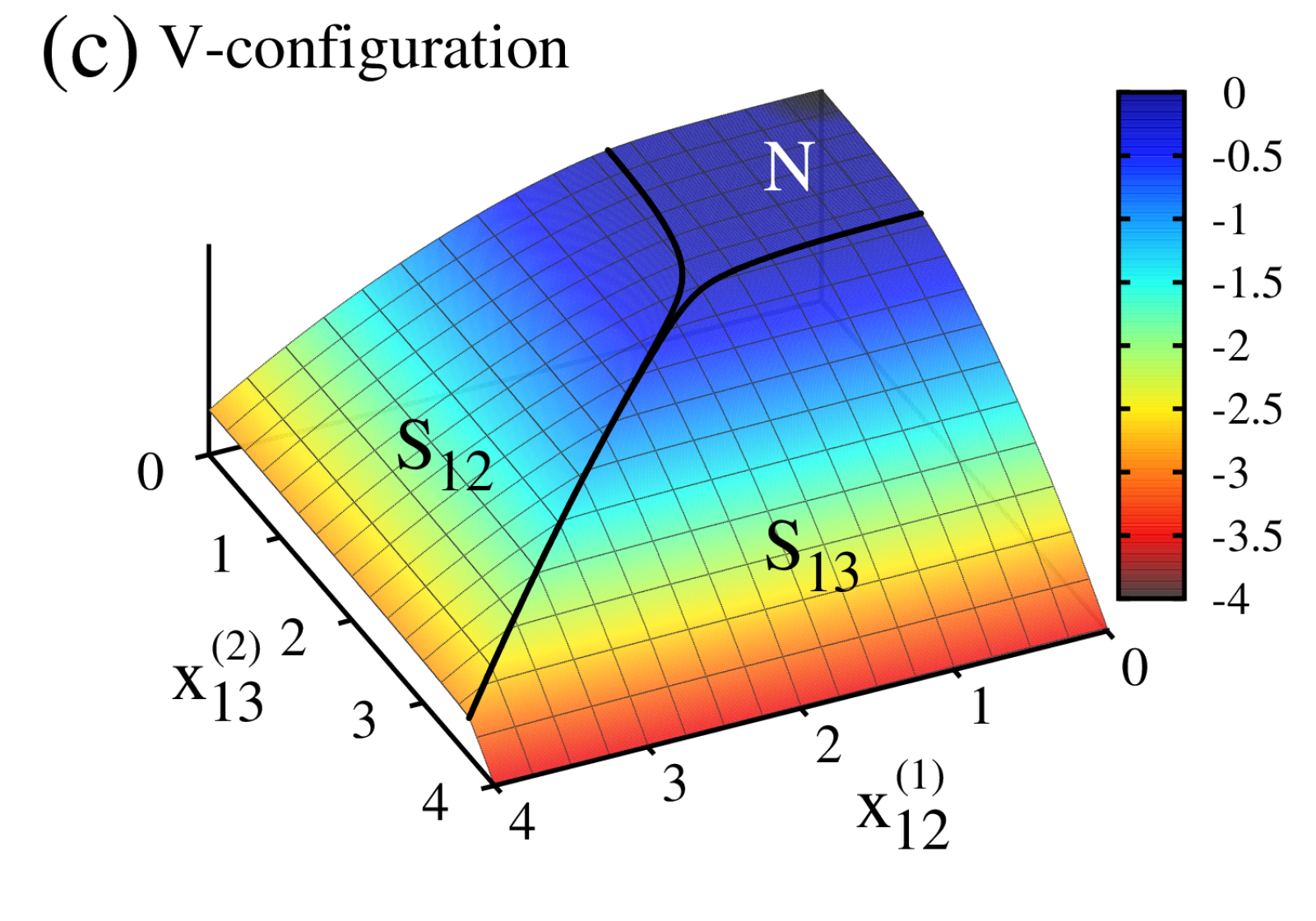}
\end{center}
\caption{(Colour online.) Ground state energy surface and quantum separatrices (lines) for $N_a=4$ particles is shown for $3$-level atoms as function of the dimensionless dipolar strengths, for the atomic configurations (a) $\Xi$, (b) $\Lambda$ and (c) $V$. The atomic configurations are: (a) $\Xi$, (b) $\Lambda$ and (c) $V$. Parameter values are given in figure~\ref{f.eminRWANa1}.}
\label{f.eminNa4}  
\end{figure}

\subsection{Generalised Dicke Model}
\label{GDM}

In a similar way to the GTCM model, figures~\ref{f.eminNa1} and \ref{f.eminNa4} show the energy surface per particle as function of the dimensionless coupling strengths for GDM model. Here the separatrix is obtained by means of the fidelity concept in quantum information.

In figure~\ref{f.eminNa1} we distinguish $3$ types of lines that run across the energy surfaces. Continuous: there is no change of parity as we cross them, and the fidelity between neighbouring states on each side remains close (but different) to $1$; these are the ones that become second-order transitions in the thermodynamic limit (i.e., they do not become discontinuous as $N_a$ increases). Dashed: there is a change of parity as we cross them; these appear for $N_a$ odd, transitions across these lines are discontinuous, and the fidelity between neighbouring states on each side falls to $0$; as $N_a$ increases, the two lines approach each other and become one in the thermodynamic limit; see also figure~\ref{f.rho337vs366} below. Dotted: there is no change of parity as we cross them; the transition is continuous (of second order) while the lines are apart, with the fidelity between neighbouring states on each side being close (but different) to $1$, so we call it a {\it continuous stable transition}; as the dotted lines approach each other the fidelity falls close (but different) to $0$, leading to a {\it continuous unstable transition}, and a first-order transition when they truly coalesce, which happens at earlier values of the coupling constants as $N_a$ increases.

In figure~\ref{f.eminNa1}(a), we show for the $\Xi$-configuration that the ground state energy surface is constituted by states with two parities: $ee$ (even-even), and $oe$ (odd-even). The normal region is delimited by the continuos transitions (solid lines) in the sectors with fixed parity $ee$ and $oe$, and by a discontinuos transition (dashed line) that runs in-between two dotted lines. Across these dotted lines in the superradiant regions $S_{12}$ and $S_{23}$, the transitions preserve parity and are continuous. For small values of $x_{12}$, the border between the normal and region $S_{23}$ presents a continuous transition (incomplete continuous trajectory), which appears incomplete due to the fact that, in the calculation of the separatrix, trajectories with fixed value of $x_{12}$ were used to evaluate the fidelity which depends on only one parameter Eq.~(\ref{eq.fide}). The quantum separatrix is formed by three kinds of lines: The dashed line separates the $ee$- and $oe$- energy surfaces, hence there are discontinuous quantum transitions along this line. The continuous line on the $ee$-energy surface provides continuous transitions from one ground state to another of the same parity. The dotted lines on the $oe$-energy surface divide solutions where one of the photon modes $\Omega_1$ or $\Omega_2$ dominates; across these lines, for small values of the control parameters where there are two lines the transitions are continuous, while where these lines coalesce one has discontinuos transitions. This is in agreement with the variational calculation~\cite{cordero15}.

For the $\Lambda$-configuration we find that the ground state energy surface is constituted by regions where the parities $eo,\, ee$, and  $oo$ are preserved~[cf. Fig.~\ref{f.eminNa1}(b)]. The normal region is divided into two sub-regions where states with parity $eo$ or $ee$ dominate. Similar to the case of the dashed lines in Fig.~\ref{f.eminNa1}(a), these divide the three regions where parity is preserved. The continuous line divides the normal and collective regions of $ee$ and $oe$ parity, while the dotted lines divide the $oo$-surface where states exhibit only one photon mode in the solution. Across these dotted lines we have continuous transitions for small values of $x_{13}$ and $x_{23}$, and discontinuous transitions for large values of the coupling parameters (where the lines coalesce). 

For the $V$-configuration one finds that the ground energy surface has only states with $ee$ parity [cf. Fig.~\ref{f.eminNa1}(c)]. The separatrix exhibits continuous transitions from the normal to the collective regions for small values of the dipolar couplings, while for large values of these the transitions from $S_{12}$ to $S_{13}$ are discontinuous (again where the dotted lines coalesce).

In contrast to the solution for $N_a=1$, we find that for $N_a=4$ the ground state energy surface has only $ee$ parity states for the three atomic configurations. These are shown in Fig.~\ref{f.eminNa4}. Their separatrices present both kinds of transition. For the $\Xi$ configuration the transition is continuous from the normal to the $S_{12}$ regions, for the $\Lambda$ case the transition is continuous from the normal to the $S_{13}$ regions, and for the $V$ configuration it is continuous from the normal to both superradiant regions. All other crossings are discontinuous.

It is important to stress that the parities refer to those of the constants of motion chosen to work with from table~\ref{t.opK}, for each atomic configuration. The parity of the ground state will also change with the number of particles $N_a$. In our case, with the parameters given in figure~\ref{f.eminRWANa1} and the symmetries given by the operators Eqs.~(\ref{eq.k1k2X}),~(\ref{eq.k1k2L})~and~(\ref{eq.k1k2V}), we find that for an even number of particles the ground state energy surface possesses an $ee$ parity for the $\Xi$- and $\Lambda$-configurations, while for an odd number of particles it presents different parities. In contrast, for the $V$-configuration the ground state energy surface is formed by states with an $ee$ parity independently of the number of particles.

\section{Quantum Phase Transitions}
\label{phase_transitions}

\begin{figure}
\begin{center}
\includegraphics[width=0.6\linewidth]{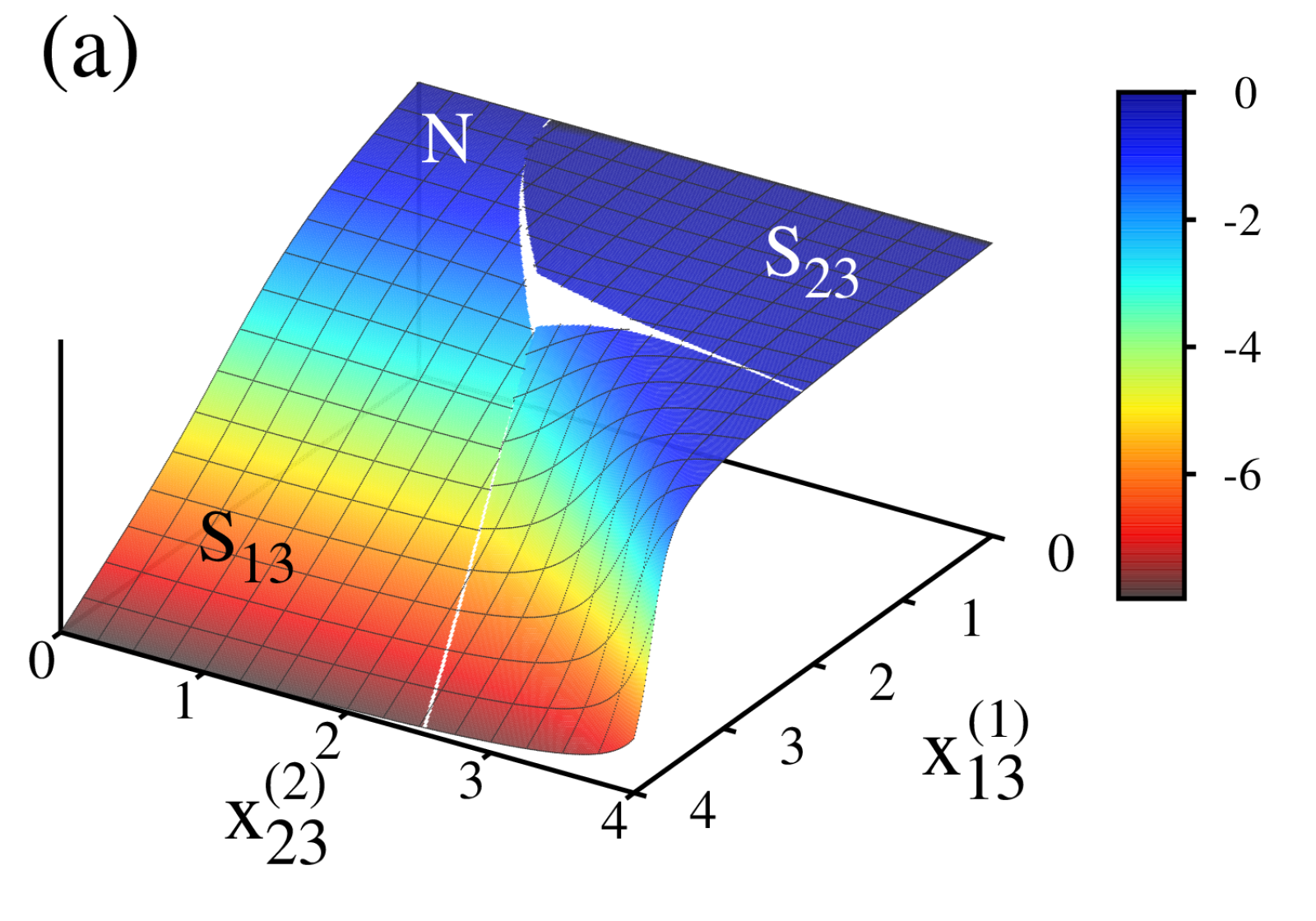}\\
\includegraphics[width=0.6\linewidth]{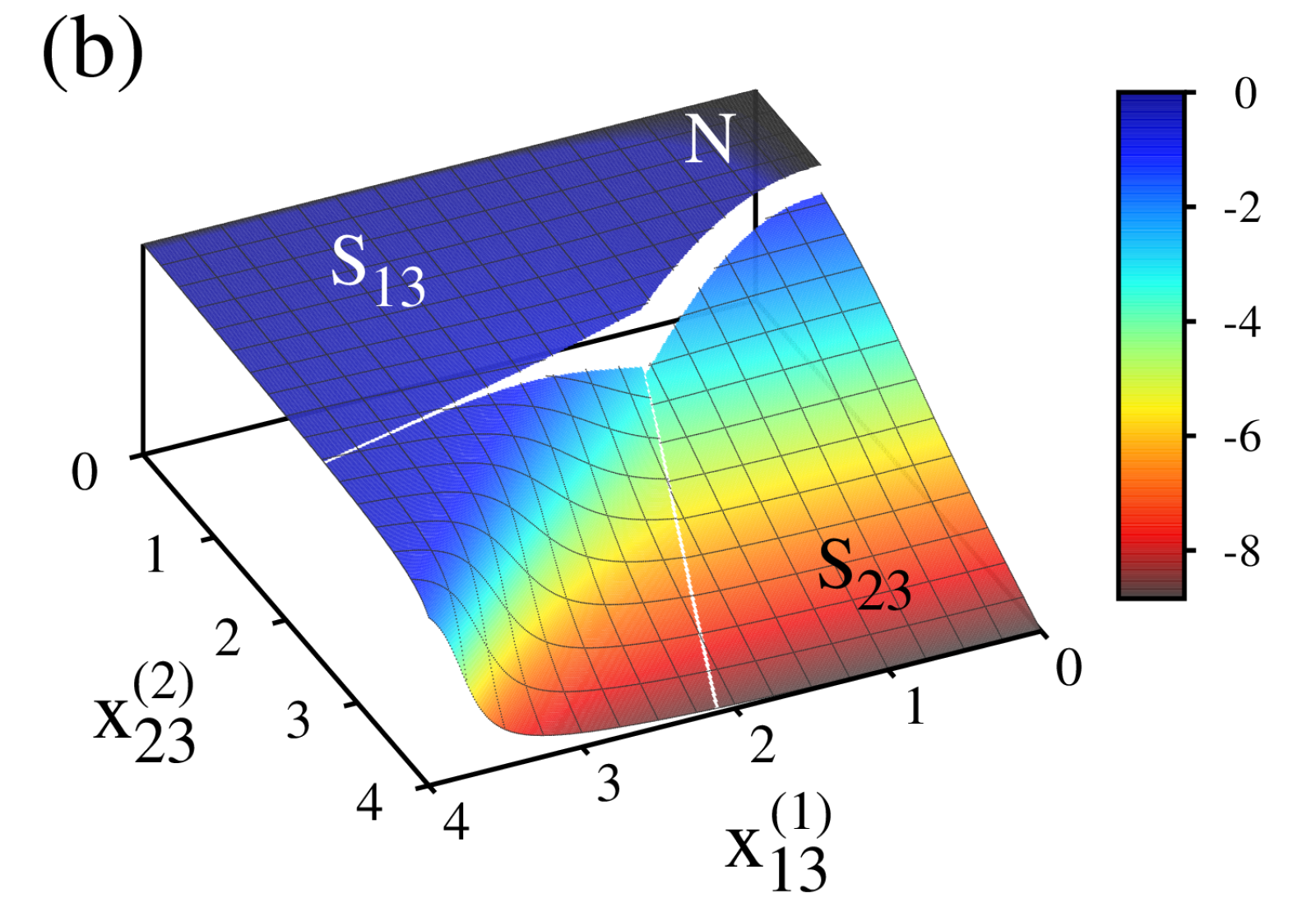}
\end{center}
\caption{(Colour online.) Derivatives of the ground state energy surface for a single atom in the $\Lambda$-configuration with respect to (a) $x_{13}^{(1)}$ and (b) $x_{23}^{(2)}$ [see Fig.~\ref{f.eminNa1}(b)].}
\label{f.dELNa1}  
\end{figure}

In this section we consider the Dicke Hamiltonian. In order to characterise the types of transitions in the quantum phase diagram, we consider, without loss of generality, the case of a single particle in the $\Lambda$-configuration, since in this case the phase diagram presents all kinds of transitions [cf. Fig.~\ref{f.eminNa1}(b)].

We use the Ehrenfest classification~\cite{gilmore93} and consider the lowest derivatives for which the ground state energy surface remains continuous, but this requires necessarily numerical calculations and these progressively loose precision, except for the first order derivative which is evaluated exactly using the Hellmann-Feynman theorem~\cite{hellmann37, feynman39, cohen77}
\begin{equation*}
\frac{{\rm d}}{{\rm d} \mu} \langle \psi|\op{H}(\mu)|\psi\ket = \left\langle \psi\left|\frac{{\rm d}}{{\rm d} \mu} \op{H}(\mu)\right|\psi\right\ket\,,
\end{equation*}
for $|\psi\ket$ an eigenstate of $\op{H}$.
In figure~\ref{f.dELNa1} the derivative with respect to the control parameters $x_{ij}$ of the ground state energy surface is shown. Discontinuities of the derivative, indicating a first order transition, appear when a change of parity occurs [see dashed lines in Fig.~\ref{f.eminNa1}(b)]; one may observe that the discontinuity vanishes as the control parameters grow. We also note that the region $S_{23}$ has a small dependence on $x_{13}$ (its corresponding derivative is close to zero) Fig.~\ref{f.dELNa1}(a), while the region $S_{13}$ has a small dependence of $x_{23}$ Fig.~\ref{f.dELNa1}(b).

The dependence on $x_{jk}$ of the energy surface, i.e., $\frac{\partial}{\partial x_{jk}} E_g\approx0$, has as a consequence that the collective region divides itself into monochromatic regions, because the corresponding dipolar interaction term does not play an important role in the contributions of the ground state and hence a sub-system with a single mode determines the bulk of the ground state. Notice also that the derivative of the ground state energy surface clearly marks the boundary between these regions.
%
\begin{figure}
\begin{center}
\includegraphics[width=0.65\linewidth]{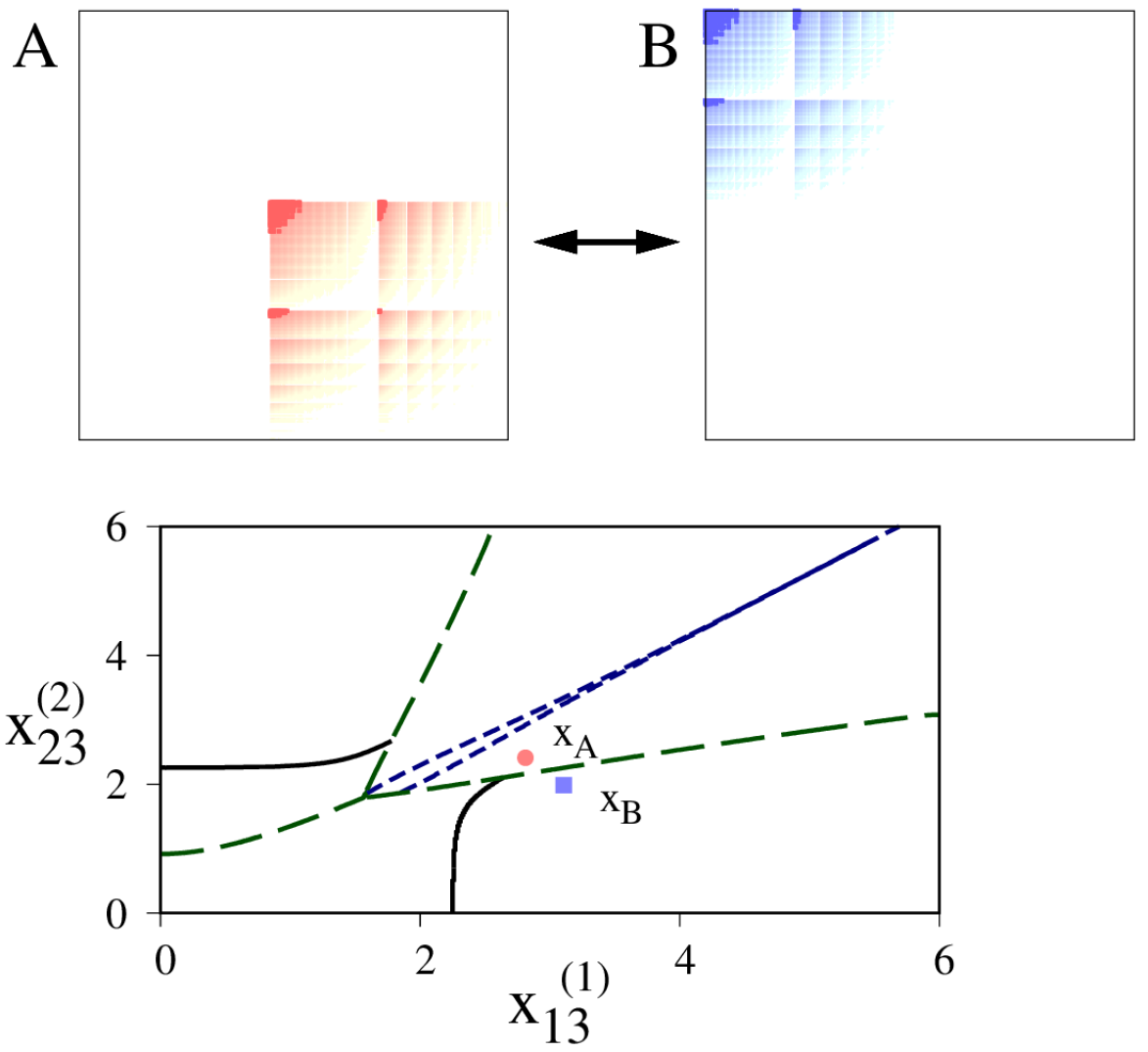}
\end{center}
\caption{(Colour online.) Above: Density matrices at points  $x_A=(2.812,2.41)$ and $x_B=(3.11,1.99)$ in the figure below, across a discontinuous transition in the space of parameters corresponding to the phase diagram of the $\Lambda$ configuration for $N_a=1$. For these regions we have ${\rm Tr}\, \op{\rho_A}\, \op{\rho_B} =0$ and $D_B = \sqrt{2}$.}
\label{f.rho337vs366}  
\end{figure}

Since the Ehrenfest classification may not permit us to characterise the kind of transition present, due to the possible loss of precision in the numerical calculations, one may catalogue them only as {\it continuous} or {\it discontinuous}, and distinguish these by using a test based on the fidelity between neighbouring states Eq.~(\ref{eq.fide}). This classification, however, determines a discontinuous transition only when $F_\delta(\xi)=0$ since this condition is met when the subspace where the ground state at point $\xi$ lies is orthogonal to the subspace where the ground state at point $\xi+\delta$ lies.

As mentioned above, there are situations with $F(\xi)\neq 0$ and it either remains different from zero as $N_a$ increases, or reaches zero in the large $N_a$ limit. In this case, a good measure of the difference between two states $\rho_A$ and $\rho_B$ across two regions is given by the Bures distance.


One may visually see this situation by plotting the density matrices of the states at two points lying on different sides of a separatrix. Figure~\ref{f.rho337vs366} gives a visual representation of the values of elements in a matrix. We show here the density matrices of ground states at the points $A$ and $B$ indicated in the bottom panel, corresponding to the phase diagram of the $\Lambda$ configuration for $N_a=1$. Across this separatrix a change of parity occurs (dashed line), and hence a discontinuous transition occurs. The height, represented by the colour intensity, is the absolute value of the probabilities and coherences in the density matrices. In this case, darker-coloured areas indicate values $|\rho_{ij}|\geq 10^{-3}$, while white regions correspond to $|\rho_{ij}|<10^{-16}$; so the coloured region provides the bulk of the ground state. The matrices take the structure of blocks, due to the ordering of the reduced bases taken~\cite{cordero19}; clearly, the matrices are orthogonal, as we have ${\rm Tr}\, \op{\rho_A}\, \op{\rho_B} =0$. This trace of the product of the density matrices at {\it A} and {\it B} is a measure of the fidelity between the states, i.e., ${\rm Tr}\, \op{\rho_A}\, \op{\rho_B} = \vert\langle\Psi_A\vert\Psi_B\rangle\vert^2$.

\begin{figure}
\begin{center}
\includegraphics[width=0.65\linewidth]{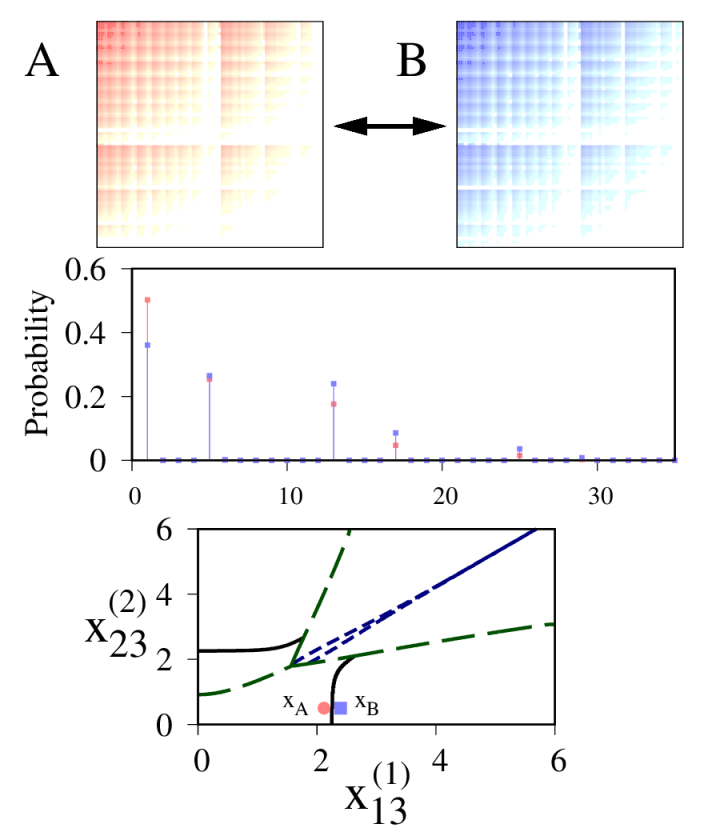}
\end{center}
\caption{(Colour online.) Above: Density matrices at points $x_A=(2.12,0.50)$	and $x_B=(2.40,0.50)$ in the figure below, across a stable-continuous transition in the space of parameters corresponding to the phase diagram of the $\Lambda$ configuration for $N_a=1$. For these states we have ${\rm Tr}\, \op{\rho_A}\, \op{\rho_B} = 0.971$ and $D_B = 0.171$. The middle figure plots the diagonal elements of the reduced density matrix for the field, at points $x_A$ and $x_B$, plotted against the basis states, for the first $35$ of $103$ states; see text for details.}
\label{f.rho70vs84}  
\end{figure}
%
\begin{figure}
\begin{center}
\includegraphics[width=0.6\linewidth]{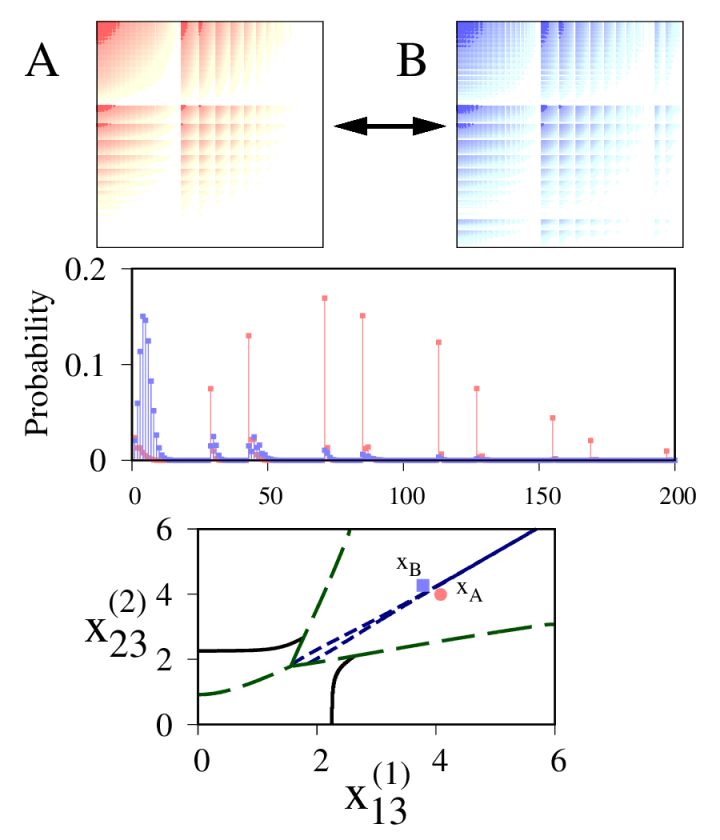}
\end{center}
\caption{(Colour online.) Above: Density matrices at points $x_A=(4.08,3.99)$  and $x_B=(3.78,4.27)$ in the figure below, across an unstable-continuous transition in the space of parameters corresponding to the phase diagram of the $\Lambda$ configuration for $N_a=1$. For these states we have ${\rm Tr}\, \op{\rho_A}\, \op{\rho_B} = 0.240$ and $D_B = 1.010$. The middle figure plots the diagonal elements of the reduced density matrix for the field, at points $x_A$ and $x_B$, plotted against the basis states for the first $200$ of $588$ states; see text for details.}
\label{f.rho461vs486}  	
\end{figure}

Now, for the case of a continuous transition with $F_\delta(\xi)\neq0$, the phase diagram of the $\Lambda$ configuration for a single particle figure~\ref{f.eminNa1}(b) shows the stable-continuous transitions indicated by continuous lines, while the unstable-continuous with short-dotted lines.

In figure~\ref{f.rho70vs84} the density matrices of two points around a stable-continuous transition are plotted. One may observe that the bulk of the ground state grows smoothly as the control parameter moves from point $A$ to point $B$; in fact for these states one finds ${\rm Tr}\, \op{\rho_A}\,\op{\rho_B} = 0.971$ and $D_B = 0.171$. The middle figure plots the diagonal elements of the reduced density matrix for the field, at points $x_A$ and $x_B$. These are the probabilities of having $\nu_1$ photons of type $1$ and $\nu_2$ photons of type $2$, plotted against the basis states, where these are ordered in the same manner as for the full density matrices. It is clear that the same basis states contribute to the photon populations, being in agreement with the high Fidelity and small Bures distance between the states.

On the other hand, figure~\ref{f.rho461vs486} shows the situation for an unstable-continuous transition: the density matrices make clear that the bulk of the ground state suffers a very significant change between neighbouring states. In this case we find ${\rm Tr}\, \op{\rho_A}\, \op{\rho_B} = 0.240$ and $D_B = 1.010$, and this is an indicator that abrupt changes occur on observables. Note that new columns at right (and corresponding rows at bottom) appear for $x_B$, denoting new states that contribute at this point. In particular, a study of the Wigner quasi-probability distribution function exhibits these abrupt changes~\cite{lopez-pena21}. Again, the middle figure plots the diagonal elements of the reduced density matrix for the field, at points $x_A$ and $x_B$. Note that the states that contribute to the photon populations are very different, again consistent with the small Fidelity and large Bures distance between the states.

\section{Reduced Density Matrix for the Matter}
\label{matter}

\begin{figure}
\begin{center}
\includegraphics[width=0.6\linewidth]{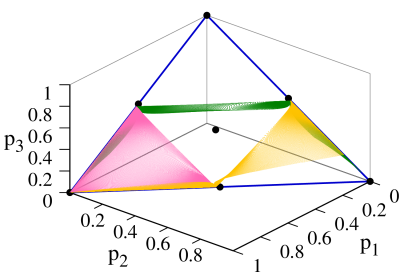}
\caption{(Colour online.) Simplex of a density matrix of $3$ dimensions. The locus of points available for the one particle case of the atomic configurations $\Xi$ (yellow), $\Lambda$ (green), and $V$ (pink) are shown. The special set of points representing pure states: $\{ (1,0,0), (0,1,0), (0,0,1)\}$ and mixed states, $\{(1/2,1/2,0),(1/2,0,1/2), (0,1/2,1/2), (1/3,1/3,1/3) \}$ are displayed as black dots.}
\label{simplex}
\end{center}
\end{figure}

In the one particle case, and for the three atomic configurations, one may show by simple inspection of the basis states, that the reduced density matrix takes a diagonal form in terms of the occupation probabilities $p_k$ of the three-level system, and can therefore be written as
\[
\rho_M =\left(
\begin{array}{ccc}
p_1  & 0  & 0  \\
0  &  p_2 & 0  \\
0  &  0 & p_3  
\end{array}
\right) \, ,
\]
where $p_3=1-p_1-p_2$. These occupation probabilities are also related to the expectation values of the matter observables $\bm{A}_{kk}$, with $k=1,2,3$, and may be measured experimentally, which allows for the possibility to check, experimentally, the theoretical phase diagram proposed. Geometrically, this type of density matrix is represented by a $2$-simplex~\cite{bengtsson17}, where the vertices denote pure states (cf.Fig.~\ref{simplex}).
Mixed density matrices are associated to the border lines, and denote the subset of systems where only two occupation probabilities are relevant, $p_1-p_2$, $p_1-p_3$, or $p_2-p_3$, while all the interior points correspond to mixed states with the three levels present.

The simplex for the reduced matrix of the matter, for the one particle systems of the generalized Dicke model, gives a visual information about the entanglement properties. If the system occupies the vertices we have a pure system without entanglement between matter and field. However all the other points indicate a mixed state. Additionally, the points along the edges of the triangle show the dominance of only two levels of the system, which is the prediction of the quantum phase transitions for the considered models in the limit $N_a \to \infty$. The variational result will thus be present only at the edges, as we vary the control parameters, indicating a monochromatic system~\cite{cordero15}.

From Fig.~\ref{simplex} one can conclude the following: For the $\Xi$ atomic configuration (yellow, in the figure) there are regions in the parameter space of the dipolar coupling strengths where only two-level subsystems are affected by the Hamiltonian, these are denoted by $S_{12}$ and $S_{23}$. Additionally it has a zone where the three levels play a fundamental role. The $\Lambda$ atomic configuration exhibits also two regions where there is a dominance of two-level subsystems, denoted by $S_{13}$ and $S_{23}$. Also, we have a rectangular zone where the three levels participate. Finally, for the $V$ atomic configuration there are also two-level subsystems, denoted by $S_{12}$ and $S_{13}$, and there is an extensive triangular region where the three levels take part in the behaviour of the atomic system. Notice that the two-level subsystems touch the border of the triangular simplex.

The simplex for {\it any two dimensional} density matrix is a segment where the extremes denote pure states while the rest designate mixed states.  According to Fig.~\ref{simplex} of the simplex for the three dimensional density matrix, we notice dominance of the $\{ (p_1, \, p_2), (p_2,\, p_3)\}$ occupation probabilities for the $\Xi$ atomic configuration, 
of $\{(p_1,\,p_3) , (p_2,\, p_3)\}$ for the $\Lambda$ atomic configuration, 
and of $\{(p_1,\, p_2),(p_1,p_3)\}$ for the $V$ configuration. To test and prove this information obtained from the simplex, we have calculated the sum of the occupation probabilities of those two-dimensional subsystems as a function of the corresponding coupling strengths between the two-mode electromagnetic field and the matter.

%
\begin{figure}
\begin{center}
\includegraphics[width=0.33\linewidth]{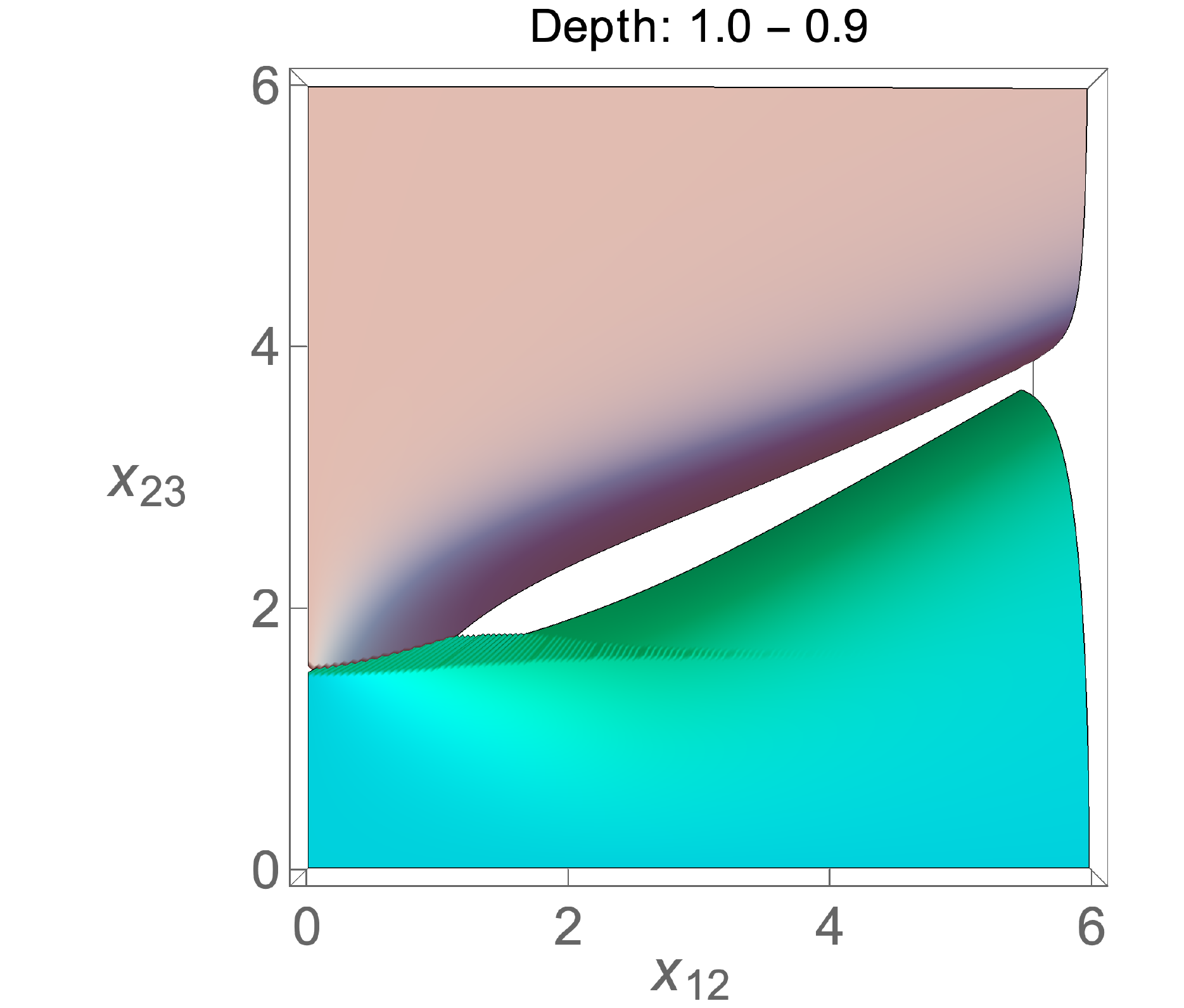}\quad
\includegraphics[width=0.295\linewidth]{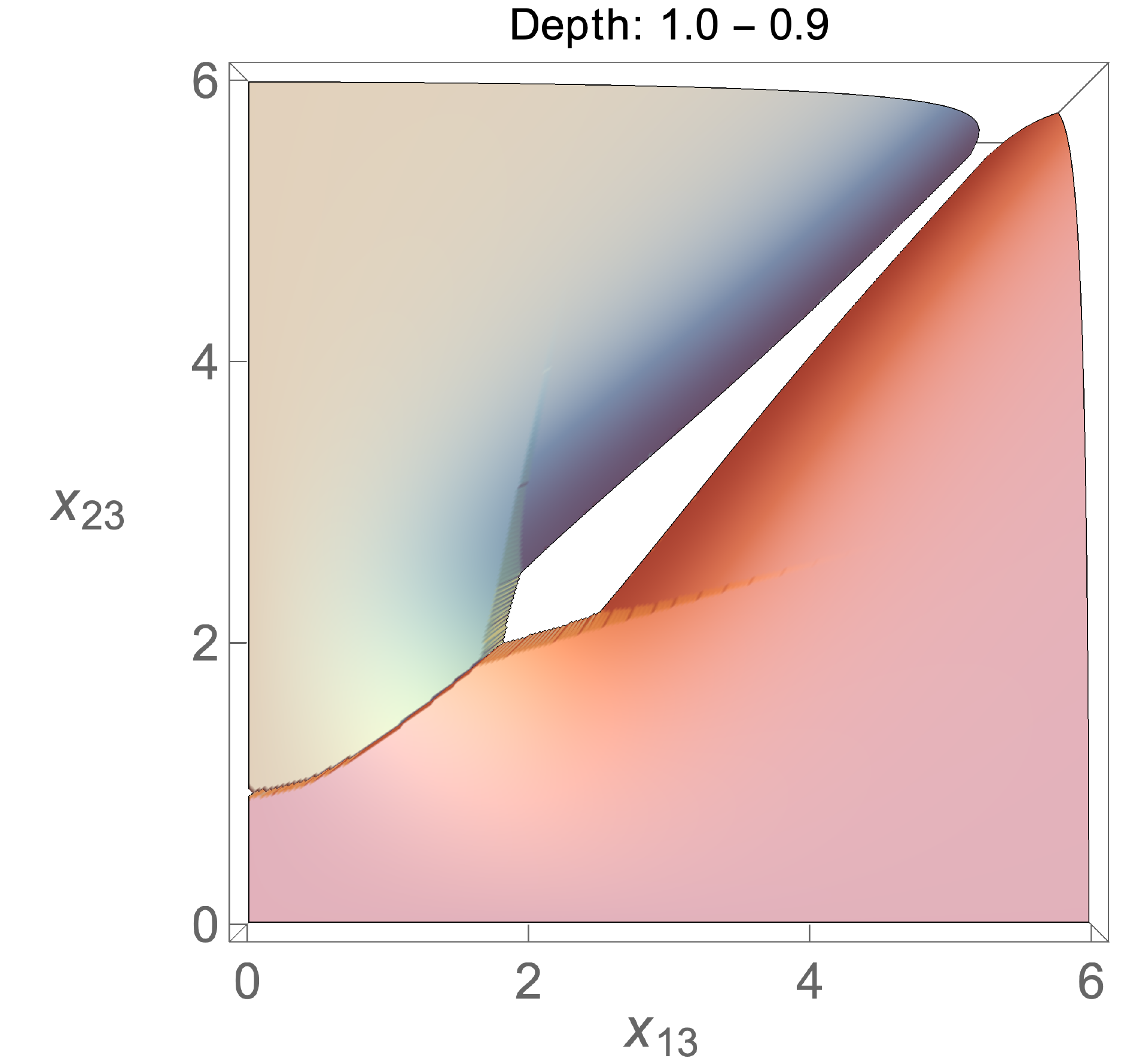}\quad
\includegraphics[width=0.3\linewidth]{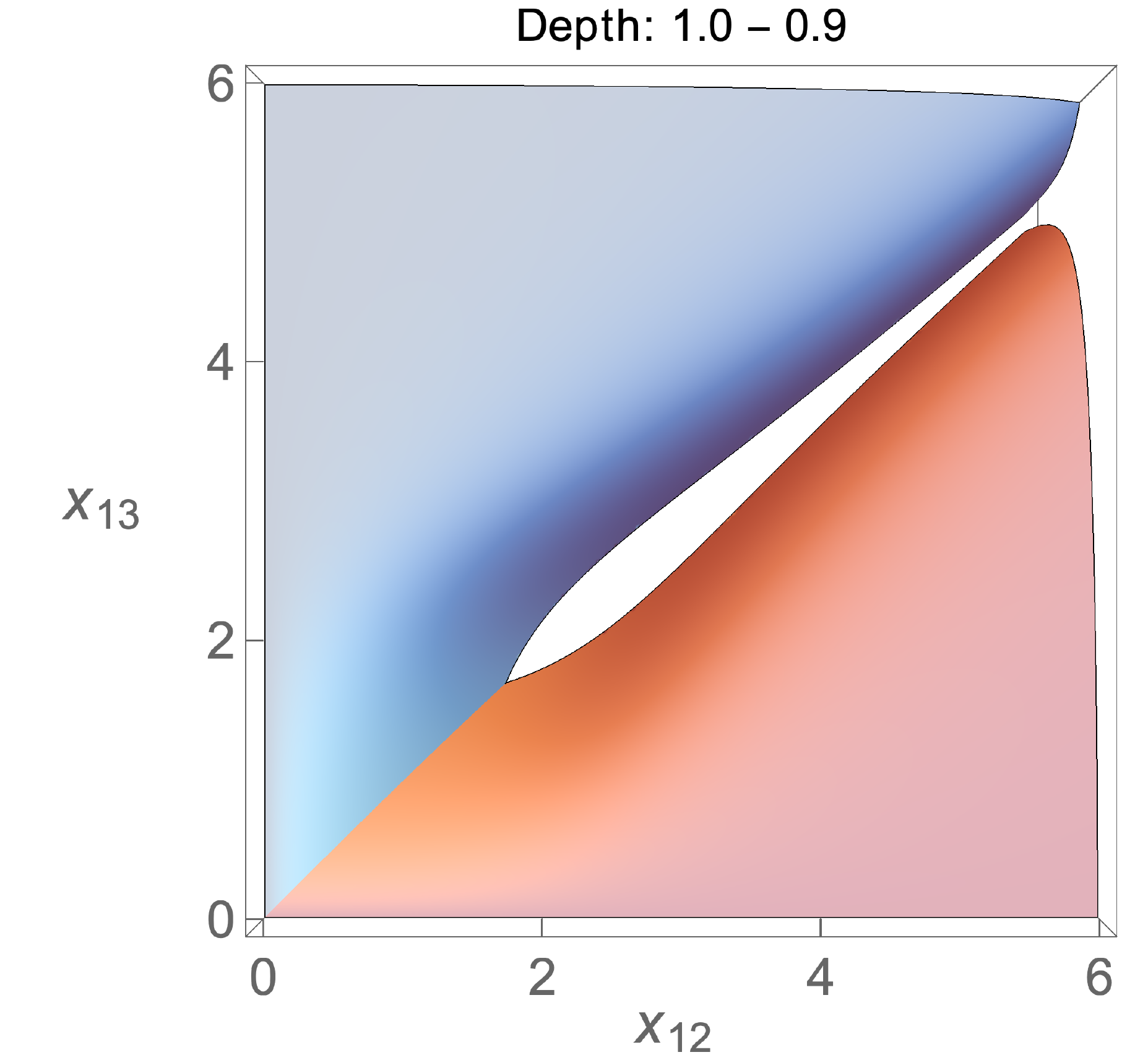}
\end{center}
\caption{(Colour online.) Left: Sum of the occupation probabilities $p_1+p_2$ and $p_2+p_3$ as a function of the matter-field interaction coupling for the $\Xi$ atomic configuration. Centre: Sum of the occupation probabilities $p_1+p_3$ and $p_2+p_3$ as a function of the matter-field interaction coupling for the $\Lambda$ atomic configuration. Right: Sum of the occupation probabilities $p_1+p_2$ and $p_1+p_3$ as a function of the matter-field interaction coupling for the $V$ atomic configuration. In all cases the depth of the surface goes from $1.0$ to $0.9$.}
\label{f.occuproba}  
\end{figure}

The results obtained are given in Fig.~\ref{f.occuproba}; at left we have plotted $p_1+p_2$ and $p_2+p_3$ as functions of $x_{12}$ and $x_{23}$ for the $\Xi$ atomic configuration; the middle graph shows the sum of the occupation probabilities $p_1+p_3$ and $p_2+p_3$ as functions of $x_{13}$ and $x_{23}$ for the $\Lambda$ configuration; and at right the corresponding sums of occupation probabilities $p_1+p_2$ and $p_1+p_3$ are exhibited as functions of the coupling strengths $x_{12}$ and $x_{13}$ for the $V$ configuration. In all cases the range of the sum of the occupation probabilities is $0.9 \leq p_i + p_j\leq 1$ with $i,j=1,2,3$.

In all three cases the results indicate a strong correspondence with the quantum phase diagrams obtained by means of the fidelity and the Bures distance concepts (see Fig.~\ref{f.eminNa1}), that is, one is able to identify the regions where the photon modes $\Omega_{12}$ and $\Omega_{23}$ dominate for the $\Xi$ case, the locus of points where the photon modes $\Omega_{13}$ and $\Omega_{23}$ dominate for the $\Lambda$ configuration, and the zones related with the dominance of the photons $\Omega_{12}$ and $\Omega_{13}$ for the $V$ atomic configuration. It is important to stress that these sums of occupation probabilities can be measured, i.e., they can be used to demonstrate the existence of the quantum phase diagrams of a finite system.

\begin{figure}[!h]
\begin{center}
{\includegraphics[width=0.6\linewidth]{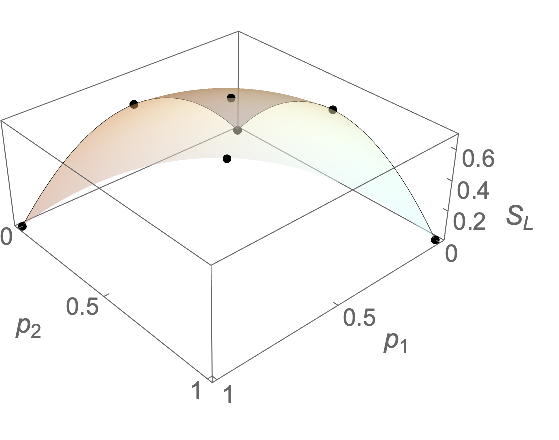}} \quad
{\includegraphics[width=0.6\linewidth]{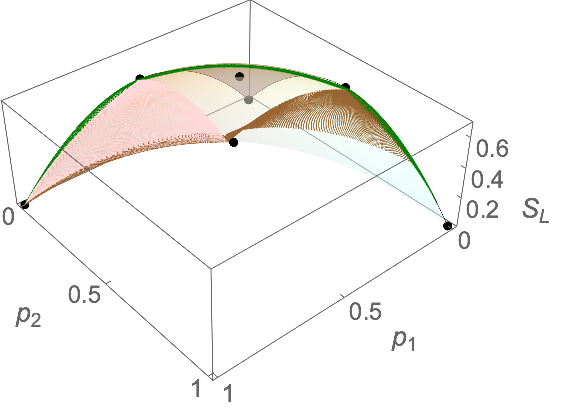}}
\caption{(Colour online.) Top: linear entropy as function of the occupation probabilities $p_1$ and $p_2$. Bottom: the corresponding values for the case of one particle for the atomic configurations $\Xi$ (brown), $\Lambda$ (green), and $V$ (pink) also shown. The special set of points representing pure states: $\{ (1,0,0), (0,1,0), (0,0,0)\}$ and mixed states, $\{(1/2,1/2,1/2),(1/2,0,1/2), (0,1/2,1/2), (1/3,1/3,2/3) \}$ are displayed as black dots.}
\label{linear}
\end{center}
\end{figure}

For this type of reduced density matrices $\rho_M$ ($3\,x\,3$ diagonal matrix) the linear entropy $S_L = 1 - {\rm Tr}(\rho^2)$ can be written in the form
\begin{equation}
S_L = 2\,  (p_1+p_2) - 2\,(p^2_1+p^2_2 + p_1\, p_2) \, . 
\end{equation}
A plot of the linear entropy is given in~Fig.\ref{linear} (top) for the $3$-dimensional density matrix (blueish brown); in~Fig.\ref{linear} (bottom) the corresponding results in the one-particle case for the atomic configurations $\Xi$, $\Lambda$ and $V$ are shown in brown, green, and pink, respectively. It is immediate that there are quantum correlations between the matter and the electromagnetic field for most values of the occupation probabilities of the energy levels of the system $1$ and $2$.

\section{Conclusions}
\label{conclusions}

We have studied finite systems, and seen that strong changes in the composition of the ground state (and even in the first excited states) take place when we cross singular regions in the energy surface. The phase diagram for a finite number of particles is much richer than that obtained in the thermodynamic limit, where the number $N_a$ of particles and the total number of excitations tend to infinity. We have therefore called these ``phase transitions'', justified by the latter argument; it is true that in the literature many authors refer to these as ``crossovers''.

In this work we have exhibited a polychromatic behaviour in the phase diagram of the quantum solution of a system of $3$-level atoms interacting with two modes of electromagnetic field, in the generalised Tavis-Cummings model (RWA approximation) and in the Dicke model. That is, the phase diagram which divides the normal and collective regions shows that the collective region divides itself into subregions where the bulk of the ground state is dominated by a sub-system with a single mode. This is in agreement with the variational solution found in~\cite{cordero15}.

For the Tavis-Cummings model we found that the polychromatic behaviour appears even for a single atom, and the phase space is formed by an infinite number of discontinuous transitions at which the constants of motion change their values Figs.~\ref{f.eminRWANa1} and \ref{f.eminRWANa4}.

Effects of the finite number of particles in the GDM model are visually exemplified by comparing the phase diagrams for $N_a=1$ and $N_a=4$ particles in Figs.~\ref{f.eminNa1} and \ref{f.eminNa4}, respectively. In fact, we found for $3$-level atoms that for an even number of particles the ground energy state does not suffer a change in parity across the phase diagram, while for an odd number of particles changes of parity appear for the $\Xi$- and $\Lambda$-configurations. We showed also that the derivative of the ground energy provides visually the collective sub-regions where a single photon contributes highly to the ground state, and that a finer characterisation of the kind of transition may be given, by using the density matrices, in terms of their continuity and stability. The same phase diagram manifests itself in the study of the Wigner function~\cite{lopez-pena21}.

When calculating the reduced density matrix for the matter sector, we see that there are regions in the parameter space of the coupling strengths where only two-level subsystems are affected by the Hamiltonian, for the three different atomic configurations. By plotting the linear entropy as a function of the occupation probabilities, it is clear that there are quantum correlations between the matter and the electromagnetic field for most values of the occupation probabilities of the energy levels of the system.

Finally, although in this work we have given numerical results for the case of $3$-level atoms, the variational result shows that this polychromatic behaviour is valid for general systems involving $n$-levels and $\ell$ modes of electromagnetic field.

\section*{Acknowledgments}
This work was partially supported by DGAPA-UNAM under projects IN101619, IN112520, and IN100120.
\vspace{0.1in}

\appendix

\section{Dimensions of the RWA basis}\label{s.rwa.dim}

In order to find the subspace degeneracy in the RWA approximation, for the different atomic configurations $\Xi$, $\Lambda$, and $V$, we  use the expression for the degeneracy of an $N$-dimensional Harmonic oscillator with $n$ quanta excitations
\begin{equation}
g_N(n):= \frac{(n+N-1)!}{n!(N-1)!}\,.
\end{equation}

This yields the following results, for fixed values of  $k_1$ and $k_2$:
%
\begin{eqnarray}
	\fl
	\resizebox{0.74\hsize}{!}{$
{\rm D}_\Xi(k_1,k_2)= \left\{ \begin{array}{ l l } 
g_3(k_1-k_2)& N_a\geq k_1-k_2\quad\&\quad k_2 \geq \displaystyle{\frac{k_1}{2}} \\[3mm] 
g_3(k_1)-g_3(k_1-k_2-1)-2\,g_3(k_2-1) & N_a\geq k_1-k_2\quad\&\quad k_2 < \displaystyle{\frac{k_1}{2}}\\[3mm]
g_3(N_a)-g_3(N_a-k_2-1)& N_a>k_2\quad \& \quad N_a< k_1-k_2\\[3mm]
g_3(N_a) & N_a\leq k_2\quad \& \quad N_a< k_1-k_2 \end{array}\right.\,,$}
\end{eqnarray}

\begin{eqnarray}
	\fl
	\resizebox{0.65\hsize}{!}{$
{\rm D}_\Lambda(k_1,k_2)= \left\{ \begin{array}{ l l } 
g_3(k_2) & k_2 < k_1 \quad \& \quad N_a\geq k_2 \\[3mm]
g_3(k_1) & k_2 \geq k_1 \quad \& \quad N_a\geq k_2 \\[3mm]
g_3(N_a+k_1-k_2) & k_2 \geq k_1 \quad \& \quad N_a < k_2 \\[3mm] 
g_3(N_a) & k_2 < k_1 \quad \& \quad N_a\leq k_1\quad \& \quad N_a < k_2 
\end{array}\right.\,,$}
\end{eqnarray}

\begin{eqnarray}
	\fl
	\resizebox{0.93\hsize}{!}{$
{\rm D}_V(k_1,k_2)= \left\{ \begin{array}{ l l } 
g_2(k_1)\,g_2(k_2) & N_a\geq k_1+k_2 \\[5mm] 
g_2(k_2)\,g_2(N_a)-g_3(k_2-1) & \left\{\begin{array}{l}\displaystyle{N_a\leq \frac{k_1+k_2}{2}\quad\&\quad k_2 < N_a} \quad || \\[3mm] \displaystyle{ k_1+k_2 > N_a > \frac{k_1+k_2}{2}\quad\&\quad N_a < k_1}\end{array}\right.\\[10mm]
g_2(k_1)\,g_2(N_a)-g_3(k_1-1) & \left\{\begin{array}{l}\displaystyle{N_a\leq \frac{k_1+k_2}{2}\quad\&\quad k_1 < N_a} \quad || \\[3mm] \displaystyle{ k_1+k_2 > N_a > \frac{k_1+k_2}{2}\quad\&\quad N_a < k_2}\end{array}\right.\\[10mm]
\left.\begin{array}{l}
1+g_2(k_1-1)\,g_2(k_2-1)+  \\[3mm]
g_2(N_a-1)\,g_2(k_1+k_2-1)-\\[3mm]g_3(k_1+k_2-2)-g_3(N_a-2)
\end{array}\right\} & \displaystyle{k_1+k_2>N_a> \frac{k_1+k_2}{2} \quad \& \quad N_a\geq k_1 \quad\&\quad N_a\geq k_2}\\[10mm]
g_3(N_a) & \displaystyle{k_2 \geq N_a +1 \quad\&\quad k_1 \geq N_a +1}
\end{array}
\right.\,.$}
\end{eqnarray}

\section*{References}


\begin{thebibliography}{10}
\expandafter\ifx\csname url\endcsname\relax
  \def\url#1{{\tt #1}}\fi
\expandafter\ifx\csname urlprefix\endcsname\relax\def\urlprefix{URL }\fi
\providecommand{\eprint}[2][]{\url{#2}}

\bibitem{sondhi97}
Sondhi S~L, Girvin S~M, Carini J~P and Shahar D 1997 {\em Rev. Mod. Phys.\/}
  {\bf 69}(1) 315--333

\bibitem{sachdev11}
Sachdev S 2011 {\em Quantum Phase Transitions\/} 2nd ed (Cambridge University
  Press)

\bibitem{hepp73}
Hepp K and Lieb E~H 1973 {\em Ann. Phys.\/} {\bf 76} 360--404

\bibitem{nussenzveig73}
Nussenzveig H~M 1973 {\em Introduction to Quantum Optics\/} (Gordon and Breach
  Science Publishers) ISBN 0 677 03900 X

\bibitem{einstein17}
Einstein A 1917 {\em Phys. Z.\/} {\bf 18} 121--128

\bibitem{yoo85}
Yoo H~I and Eberly J~H 1985 {\em Phys. Rep.\/} {\bf 118} 239--337

\bibitem{baksic13}
Baksic A, Nataf P and Ciuti C 2013 {\em Phys. Rev. A\/} {\bf 87}(2) 023813

\bibitem{liu17}
Liu M, Chesi S, Ying Z~J, Chen X, Luo H~G and Lin H~Q 2017 {\em Phys. Rev.
  Lett.\/} {\bf 119}(22) 220601

\bibitem{wei18}
Wei B~B and Lv X~C 2018 {\em Phys. Rev. A\/} {\bf 97}(1) 013845

\bibitem{braak16}
Braak D, Chen Q-H, Murray T~B and Solano E 2016 {\em J. Phys. A} {\bf 49} (30) 300301


\bibitem{liberti10}
Liberti G, Piperno F and Plastina F 2010 {\em Phys. Rev. A\/} {\bf 81}(1)
  013818

\bibitem{zanardi06}
Zanardi P and Paunkovi\ifmmode~\acute{c}\else \'{c}\fi{} N 2006 {\em Phys. Rev.
  E\/} {\bf 74}(3) 031123

\bibitem{vieira10}
Vieira V~R 2010 {\em J. Phys.: Conf. Ser,\/} {\bf 213} 012005

\bibitem{gu10}
Gu S~J 2010 {\em Int. J. Mod. Phys. B\/} {\bf 24} 4371--4458

\bibitem{you07}
You W~L, Li Y~W and Gu S~J 2007 {\em Phys. Rev. E\/} {\bf 76}(2) 022101

\bibitem{wang18}
Wang Y, You W~L, Liu M, Dong Y~L, Luo H~G, Romero G and You J~Q 2018 {\em New
  Journal of Physics\/} {\bf 20} 053061

\bibitem{larson17}
Larson J and Irish E~K 2017 {\em J. Phys. A: Math. Theor.} {\bf 50} 174002

\bibitem{castanos09b}
Casta\~nos O, Nahmad-Achar E, L\'opez-Pe\~na R and Hirsch J~G 2009 {\em Phys. Scr.} {\bf 80} 055401

\bibitem{castanos11}
Casta\~nos O, Nahmad-Achar E, L\'opez-Pe\~na R and Hirsch J~G 2011 {\em Phys. Rev. A} {\bf 84} 013819

\bibitem{cordero19}
Cordero S, Casta\~nos O, L\'opez-Pe\~na R and Nahmad-Achar E 2019 {\em Phys.
  Rev. A\/} {\bf 99}(3) 033811

\bibitem{cordero19b}
Cordero S, Nahmad-Achar E, Casta\~nos O and L\'opez-Pe\~na R 2019 {\em Phys.
  Rev. A\/} {\bf 100}(5) 053810

\bibitem{cordero16}
Cordero S, Casta\~nos O, L\'opez-Pe\~na R and Nahmad-Achar E 2016 {\em Phys.
  Rev. A\/} {\bf 94}(1) 013802

\bibitem{cordero15}
Cordero S, Nahmad-Achar E, L\'opez-Pe\~na R and Casta\~nos O 2015 {\em Phys.
  Rev. A\/} {\bf 92}(5) 053843

\bibitem{cordero13a}
Cordero S, {L\'opez-Pe\~na} R, {Casta\~nos} O and {Nahmad-Achar} E 2013 {\em
  Phys. Rev. A\/} {\bf 87}(2) 023805

\bibitem{cordero13b}
Cordero S, {Casta\~nos} O, {L\'opez-Pe\~na} R and {Nahmad-Achar} E 2013 {\em J.
  Phys. A: Math. Theor.\/} {\bf 46} 505302

\bibitem{nahmad-achar14}
Nahmad-Achar E, Cordero S, {Casta\~nos} O and {L\'opez-Pe\~na} R 2014 {\em
  Phys. Scr.\/} {\bf 2014} 014033

\bibitem{bures69}
Bures D 1969 {\em Trans. Amer. Math. Soc.\/} {\bf 135} 199--212

\bibitem{helstrom67}
Helstrom C 1967 {\em Phys. Lett. A\/} {\bf 25} 101--102 ISSN 0375-9601

\bibitem{gilmore93}
Gilmore R 1993 {\em Catastrophe Theory for Scientists and Engineers\/} (Dover)

\bibitem{hellmann37}
Hellmann H Leipzig 1937 {\em Einf\"uhrung in die Quantenchemie\/} (Franz
  Deuticke)

\bibitem{feynman39}
Feynman R~P 1939 {\em Phys. Rev.\/} {\bf 56}(4) 340--343

\bibitem{cohen77}
{Cohen-Tannoudji} C, Diu B and Lalo\"e F 1977, New York {\em Quantum
  Mechanics\/} vol~II (John Wiley \& Sons)

\bibitem{lopez-pena21}
{L\'opez-Pe\~na} R, Cordero S, Nahmad-Achar E and {Casta\~nos} O {\it Quantum
  Phase Diagrams of Matter-Field Hamiltonians II: Wigner Function Analysis} (submitted).

\bibitem{bengtsson17}
Bengtsson I and Zyczkowski K 2017 {\em Geometry of Quantum States: An
  Introduction to Quantum Entanglement\/} (Cambridge University Press; 2nd
  Edition)

\end{thebibliography}

\providecommand{\newblock}{}

\end{document}